\documentclass[ 
    aps,
    pra,
    twocolumn,
    letterpaper,
    10pt,
    superscriptaddress,
    showpacs,
    showkeys,
    notitlepage,
    amsmath, 
    amssymb, 
    floatfix 
]{revtex4} 
 
\usepackage{graphicx}
\usepackage{dcolumn}
\usepackage{bm}
\usepackage{color}
\usepackage{amssymb}
\usepackage{amsmath}
\usepackage{hyperref}

\newcommand{\bel}{\begin{equation}}
\newcommand{\eel}{\end{equation}}
\newcommand{\be}{\begin{equation*}}
\newcommand{\ee}{\end{equation*}}
\newcommand{\bal}{\begin{eqnarray}}
\newcommand{\eal}{\end{eqnarray}}
\newcommand{\ba}{\begin{eqnarray*}}
\newcommand{\ea}{\end{eqnarray*}}

\newcommand{\refeq}[1]{Eq.~(\ref{#1})}
\newcommand{\reffig}[1]{Fig.~\ref{#1}}

\newcommand{\ev}[1]{\langle #1 \rangle}
\newcommand{\ket}[1]{| #1 \rangle}
\newcommand{\bra}[1]{\langle #1 |}

\newcommand{\PP}{\mathcal{P}}

\renewcommand{\d}[1]{\!d#1\,}

\begin{document} 

\title{
Heisenberg-limited atom clocks based on entangled qubits
}
 
\author{E. M. Kessler}
\thanks{These authors contributed equally to this work}
\affiliation{Physics Department, Harvard University, Cambridge,
MA 02138, USA}
\affiliation{ITAMP, Harvard-Smithsonian Center for Astrophysics, Cambridge, MA 02138, USA}

\author{P. K\'{o}m\'{a}r}
\thanks{These authors contributed equally to this work}
\affiliation{Physics Department, Harvard University, Cambridge,
MA 02138, USA}

\author{M. Bishof}
\affiliation{JILA, National Institute of Standards and Technology, Department of Physics,
University of Colorado, Boulder, Colorado 80309-0440, USA}

\author{L. Jiang}
\affiliation{Department of Applied Physics, Yale University New Haven, CT
06520, USA}

\author{A. S. S{\o}rensen}
\affiliation{QUANTOP, Danish National Research Foundation Centre of Quantum Optics, Niels Bohr Institute,
DK-2100 Copenhagen, Denmark}

\author{J. Ye}
\affiliation{JILA, National Institute of Standards and Technology, Department of Physics,
University of Colorado, Boulder, Colorado 80309-0440, USA}

\author{M. D. Lukin}
\affiliation{Physics Department, Harvard University, Cambridge,
MA 02138, USA}


\date{\today}

\begin{abstract} 
We present a quantum-enhanced atomic clock protocol based on groups of sequentially larger
Greenberger-Horne-Zeilinger (GHZ) states,  
that achieves the best clock stability
allowed by quantum theory up to a logarithmic correction.
The simultaneous interrogation of the laser phase with
such a cascade of GHZ states realizes an incoherent version of the phase estimation algorithm that enables
Heisenberg-limited operation while extending the Ramsey interrogation time beyond the laser noise limit. 
We compare the new protocol with state of
the art interrogation schemes, and show that entanglement allow a significant quantum gain in the stability for short averaging time.
\end{abstract}

\maketitle
 

%

High precision atomic frequency standards form a cornerstone
of precision metrology,
and are of great importance for science and technology in modern society.
Currently, atomic clocks based on optical transitions achieve the most
precise \cite{Nicholson2012, Lemke2009} and accurate \cite{Chou2010,
Bloom2013} frequency references.
Additionally, the development of optical frequency combs
\cite{Eckstein1978, Reichert2000, Jones2000, Ye2003} -- establishing a coherent
link between the optical and radio frequencies -- enabled the application
of optical frequency standards to a wide range of scientific and technological
fields including astronomy, molecular spectroscopy and global
positioning systems (GPS).


The improvement of frequency standards using quantum resources, such as
entanglement \cite{Buzek1999, Andre2004,LouchetChauvet:2010fs,Rosenband2012_numerical,
Borregaard2013_nearHeisenberg},
has been actively explored in recent years.
 A characterization of  the improvement obtainable by using entanglement
 requires a detailed investigation of the decoherence present in the system.
 Previous studies have focused on two kind of noise sources: i) single particle
 decoherence resulting from the interaction of the atoms with the environment
 and ii) fluctuation in the local oscillator (LO) used to interrogate the atoms.
  It is well know that fully entangled states (e.g., Greenberger-Horne-Zeilinger
 (GHZ) states) allow for improved spectroscopic sensitivity, but in the same way
 that these states benefit from their increased sensitivity in the laser
 interrogation, they are generically prone to various types of noise sources
 canceling any quantum gain. It has therefore been long believed that such
 states fail to increase clock stability in the standard Ramsey type protocol
 regardless of the noise model being used \cite{Bollinger1996, Wineland1998,
Rosenband2012_numerical,Huelga1997}. On the other hand, it has been shown that
for clocks with local oscillator (LO) noise limited stability, the use of
moderately squeezed atomic states can yield a modest improvement over the
standard quantum limit (SQL) \cite{Andre2004,LouchetChauvet:2010fs}.
 A
recent study demonstrated further that, in principle, highly squeezed states
could achieve Heisenberg-limited stability using a complex adaptive measurement
scheme \cite{Borregaard2013_nearHeisenberg}.
 At the same time, it has been shown that the single particle
 decoherence-limited regime can be reached for long averaging time at
a logarithmic cost in the number of qubits by interrogating uncorrelated atomic
ensembles for suitably chosen times \cite{Rosenband2013, Borregaard2013}.

\begin{figure}
\centering 
\includegraphics[width=0.45\textwidth]{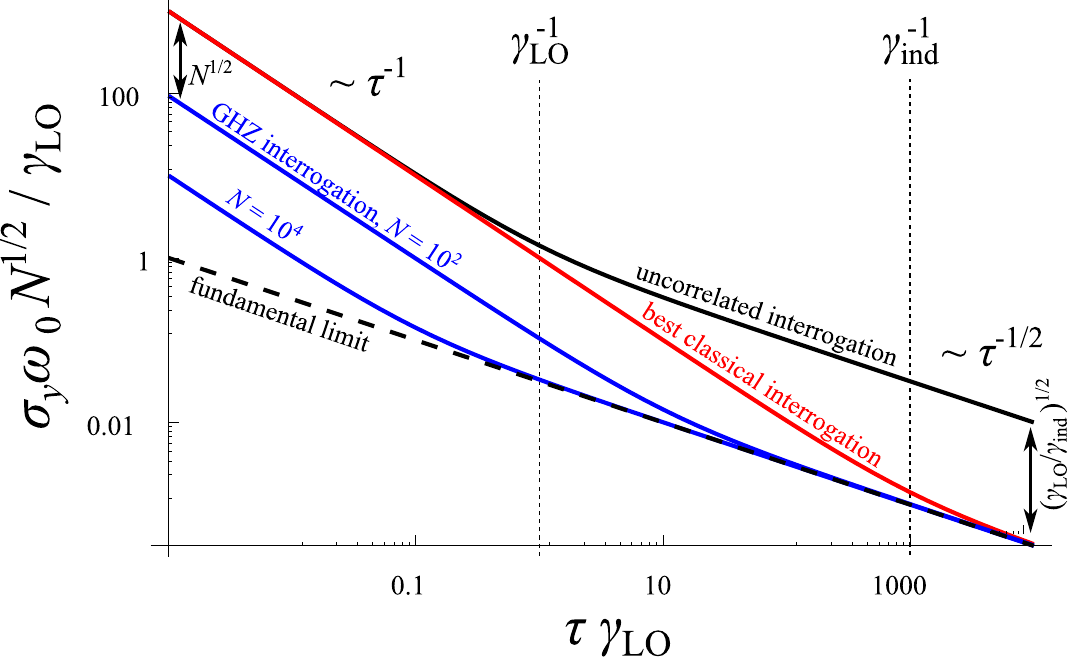}
\caption{
\label{fig:sigma_tau}
Allan deviation $\sigma_y$ for different protocols as a function of averaging
time $\tau$, normalized to the standard quantum limit, for $\gamma_\text{LO} /
\gamma_\text{ind} = 10^3$. The solid black line corresponds to the standard scheme using
a single uncorrelated ensemble. It fails to reach the fundamental noise floor
set by the atomic transition linewidth (cf. \refeq{eq:sigma(2)}, broken line).
A more sophisticated classical scheme which uses exponentially increasing Ramsey
times in each cycle \cite{Rosenband2013, Borregaard2013} allows to extend the
regime of linear scaling with $1/\tau$ up to the point where the bound
(\ref{eq:sigma(2)}) is met. In comparison, the proposed cascaded GHZ protocol
(blue solid curves) enables an $\sim N$ times faster convergence. For short
averaging times the stability is enhanced by a factor $\sqrt{N}$ as compared to
classical protocols.
}
\end{figure}

In this Letter, we  introduce a protocol involving groups of sequentially larger
GHZ states to estimate local oscillator deviations from the atomic reference in
a manner reminiscent of the phase estimation algorithm \cite{Nielsen_Chuang}.
Furthermore, we unify previous treatments of decoherence for atomic clocks 
and incorporate previous proposals involving uncorrelated atoms to
effectively narrow the LO linewidth \cite{Rosenband2013, Borregaard2013} and thereby identify
 ultimate limits to the stability of atomic clocks based on entangled atoms. The
 central results of our work are illustrated in \reffig{fig:sigma_tau}, which compares the performance of the
proposed protocol with other known approaches as a function of averaging time.
 Specifically, for LO-noise limited clocks, the
proposed quantum protocol is found to be nearly optimal, realizing the
Heisenberg limit of clock stability up to a logarithmic correction in the
particle number.
Importantly, it reaches the fundamental noise floor (dashed line) resulting from
individual dephasing of the clock qubits $N$ times faster than the best known
classical schemes, where $N$ is the total number of particles employed.

The central idea of our approach can be understood as follows.
In modern atomic clocks, the frequency of a LO is locked to an ultra-narrow
optical (or radio-frequency) transition of the clock atoms serving as the
frequency reference.
The long-term stability  of such a clock after a given total averaging time
$\tau$ is directly related to the precision by which the accumulated laser phase
relative to the atoms can be determined. To this end, in the standard Ramsey
protocol, the phase is repeatedly measured over cycle times $T<\tau$, followed
by a correction of the laser frequency according to the measurement outcome.
Since each Ramsey sequence introduces measurement noise, it is optimal to extend
the Ramsey time $T$ as much as possible, ideally to its maximum value
$T\rightarrow\tau$.
A single GHZ state consisting of $N$ entangled atoms -- whose state after the
interrogation is $\ket{\text{GHZ}}_T \propto \ket{0}^{\otimes N} + \text{exp}(-i
N\Phi_\text{LO}) \ket{1}^{\otimes N}$ -- accumulates the laser phase (denoted by
$\Phi_{LO}$) $N$ times faster than an uncorrelated state.
Thus, in principle, it allows to measure the phase with the best precision
allowed by quantum mechanics \cite{Giovanetti2011}. However, fluctuations in the
laser frequency account for the fact that the laser phase we aim to measure is a
random variable with a probability distribution that grows in width as we
increase  the Ramsey time $T$.
Whenever the laser phase realized in a particular Ramsey cycle induces a full
phase wrap on the state [i.e., the atomic phase $N \Phi_{LO}\notin[-\pi,\pi)$] a
subsequent measurement yields a $2\pi$ error in the estimation of the true laser
phase. For a single GHZ state this accounts for a strict limitation on the
maximally allowed Ramsey time in order  to limit the initial variance of
$\Phi_\text{LO}$, and the resulting laser stability is found to yield no
improvement over classical protocols  \cite{Wineland1998}.

\begin{figure}
\centering
\includegraphics[width=0.45\textwidth]{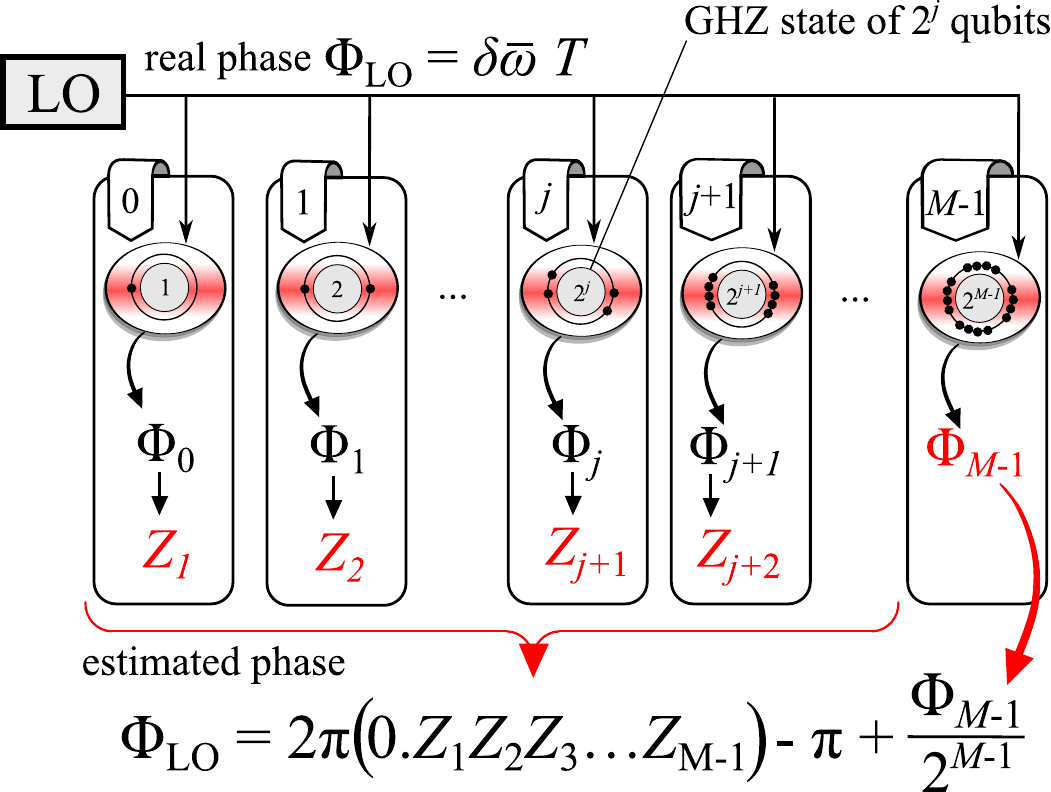}
\caption{
\label{fig:phase_estimation}
The proposed clock operation scheme employs $M$ different groups of clock atoms prepared in correlated states of varying size to interrogate the relative phase $\Phi_\text{LO}$ of the LO field.
A single group $j$ contains $n_0$ independent instances of GHZ-like states, each
entangling $2^j$ qubits, and therefore accumulating a phase $\Phi_j = 2^j
\Phi_\text{LO} \mod [-\pi,\pi]$ during a single cycle. Each group is then
used to measure this phase, which gives a direct estimate on the digit
$Z_{j+1}$ in a binary representation of the LO phase $(\Phi_\text{LO}+\pi)/2\pi=(0.Z_1Z_2Z_3\hdots)$.
This estimate is subsequently used to feedback the LO frequency. This clock operation protocol achieves the best stability allowed by quantum mechanics up to a logarithmic correction.
}
\end{figure}

To address this problem, we use a protocol reminiscent of the \textit{phase
estimation algorithm} \cite{Nielsen_Chuang} that allows a direct assessment of
the number of $2\pi$ phase wraps during the interrogation, thus enabling 
the Ramsey time to extend to its maximum value to  guarantee optimal laser
stability.
Let us assume for the moment that the accumulated laser phase after the
interrogation time $T$ lies in the interval $\Phi_\text{LO}\in[-\pi,\pi)$, and
has an exact binary representation $(\Phi_\text{LO}+\pi)/2\pi= \sum_{j=1}^M Z_j/
2^{j}$, with digits $Z_j\in \{0,1\}$ (both conditions will be relaxed below).
One can then readily show, that a GHZ state consisting of $2^{M-1}$ atoms picks
up the phase $\Phi_{M-1} = 2^{M-1} \Phi_\text{LO}  ~\text{mod}~ [-\pi,\pi) = \pi
(Z_M-1)$. Thus, by measuring if the phase is $0$ or $\pi$, the last digit of the
laser phase can be determined. However, as stated above, without knowledge of
the remaining digits (i.e., the number of phase wraps) this information is
useless.
In our protocol, these digits are found by an additional simultaneous
interrogation with successively smaller GHZ states of $2^{M-2},2^{M-3},\hdots$
entangled atoms (see \reffig{fig:phase_estimation}). Each of these states
picks up a phase proportional to its size  $\Phi_{j} = 2^{j}
\Phi_\text{LO}~\text{mod}~ [-\pi,\pi)$, and in analogy to the first state this
phase gets a contribution of $\pi (Z_j-1)$.
By distinguishing whether the phase is shifted by $\pi$ or not, we can thus
determine the value of the bit $Z_j$.
The combined information provides an estimate with an accuracy given by the
largest GHZ state, while the cascade increases the total number of atoms
employed only modestly by a factor of two: $\sum_{j=0}^{M-1} 2^j \approx
2^{M}=2\times2^{M-1}$.

However, in the limit of large averaging times, the assumption
$\Phi_\text{LO}\in [-\pi,\pi)$ is not justified anymore. Here, the optimal
Ramsey time $T\sim\tau$ can attain values that induce phase wraps of the laser
itself, causing the binary representation of the laser phase to contains digits
$Z_j\neq0$ for $j\leq0$ which are inaccessible to the technique discussed so
far.
To achieve the optimal laser stability in this regime, we extend the cascade to
the classical domain, and employ additional groups of uncorrelated atoms that
interrogate the laser with successively decreasing interrogation times, or
alternatively, using dynamical decoupling techniques \cite{Rosenband2013,
Borregaard2013,ddc}. Each of these ensembles acquires a phase that again is
reduced by multiples of two from the laser phase, and thus, following the
arguments from above, allows one to gain information on the lower digits
$Z_j$ with $j\leq0$.
The information of all digits combined provides the total number of phase wraps,
which in turn yields a Heisenberg-limited estimate of the laser phase.
By this, the protocol effectively eliminates all limitations
arising from the LO noise, and allows the Ramsey time to extend to its optimal
value to achieve the best laser stability allowed by quantum mechanics (up to a
logarithmic correction as shown below).



In the following, we provide a detailed derivation of the above results.
Modern clocks periodically measure the fluctuating LO frequency $\omega(t)$
against the frequency standard $\omega_0$ of a given ensemble of clock atoms
(qubits) to obtain an error signal.
After each Ramsey cycle of duration $T$ [i.e., at times $t_k=kT$
($k=1,2\hdots$)], the measurement data yield
 an estimate of the relative phase $\Phi_\text{LO}(t_k) =
\int_{t_k-T}^{t_k}\d{t}[\omega(t) - \omega_0]$ accumulated by the LO.
This estimate in turn is used to readjust the frequency of the LO:  $\omega(t_k)
\rightarrow \omega(t_k) - \alpha\Phi^\text{est}_\text{LO}(t_k)/T$, where
$\Phi^\text{est}_\text{LO}(t_k)$ represents a suited estimator of the phase
$\Phi_\text{LO}(t_k)$ \cite{fn1}, and $\alpha < 1$ is an suitably chosen gain.

The stability of the actively stabilized LO after a total averaging time $\tau$
is characterized by the Allan deviation (ADEV) which is directly proportional to
the measurement uncertainty $\Delta\Phi_\text{LO}(t_k)$ after each Ramsey cycle
\cite{SI},
\bel
	\label{eq:Allan-variance}
	\sigma_y(\tau) \equiv \frac{1}{\omega_0 \tau}
	\sqrt{\sum_{k = 1}^{\tau/T}\sum_{l = 1}^{\tau/T}
	T^2\ev{\delta\bar\omega_k\delta\bar\omega_l} }
	\approx
	 \frac{1}{\omega_0\sqrt{\tau T}} [\Delta\Phi_\text{LO}(T)].
\eel
Here, $\delta\bar\omega_k = \Phi_\text{LO}(t_k)/T$ is the average detuning of the
(stabilized) LO during the $k$th cycle. {%
 To obtain \refeq{eq:Allan-variance} we use the fact that after the frequency
 feedback the detuning averages become approximately uncorrelated for realistic
 laser spectra, $\ev{\delta\bar\omega_k \delta\bar\omega_l} \approx
 \ev{\delta\bar\omega^2}\delta_{kl}$ \cite{Borregaard2013,
 Andre2005,Bloom2013}}.
Other noise sources (such as the bias of the linear estimator, the Dick effect, or a sub-optimal
gain $\alpha$ \cite{Santarelli1998}) are not fundamental, and neglected in the following.


The ultimate precision by which the accumulated Ramsey
phase after each cycle can be estimated is in principle (i.e., in the limit of small phases) determined by the Cram\'{e}r-Rao
bound \cite{Rao1945,Giovanetti2011} which, e.g., is achieved by the use of GHZ states. 
However, as a consequence of the LO frequency fluctuations, in general, large phases can occasionally be acquired,  leading to uncontrolled phase wraps of the atomic phase, $\Phi(t_k)=N
\Phi_\text{LO}(t_k)\notin[-\pi,\pi)$. 
To suppress these events, the cycle time $T$ has to be
chosen such that the prior distribution of $\Phi$ is well localized
within $[-\pi,\pi)$.
This limits the maximally allowed Ramsey time to a value 
$T_\text{max} \sim\gamma_\text{LO}^{-1}/N^2$, where we assumed a white frequency noise spectrum of the LO, $S_\omega(f) =
\gamma_\text{LO}$ (for $1/f$-noise one finds the less stringent condition $T_\text{max} \sim\gamma_\text{LO}^{-1}/N$). In most cases, this value lies below the optimal (i.e., maximal) value implied by \refeq{eq:Allan-variance} $T\sim\tau$, resulting in a laser stability for GHZ states which shows no improvement over the stability achieved with uncorrelated atoms \cite{Wineland1998, Rosenband2012_numerical}.

However, unlike the individual particle noise resulting in the finite atom linewidth
$\gamma_\text{ind}$, the LO frequency fluctuations affect all clock atoms alike, and this
\textit{collective noise} does not represent a fundamental metrological limitation.
Assuming for the moment that the prior distribution of the LO phase $\Phi_\text{LO}$ after the
Ramsey time is localized in $[-\pi,\pi)$ (we will relax this condition below), we can use groups of GHZ states of varying size to measure the $\Phi_\text{LO}$ in a binary representation, as discussed above. 
In general, however, the phase does not have an exact binary representation ending at the digit $Z_{M}$. We therefore employ $n_0$ 
duplicates of the GHZ states at each level of the cascade ($n_0 =N/\sum_{j=0}^{M-1} 2^j \approx N/2^M$) to improve the precision. 
 In the case where all digits $Z_j$ ($j=1\dots, M-1$) are determined correctly according to the relation 
\bel
Z_j =
[2(\Phi_{j-1}+\pi) - (\Phi_j + \pi)]/2\pi,
\eel
the last group ($j=M-1$) then yields a Heisenberg-limited estimate of
the LO phase with accuracy $(\Delta\Phi_\text{LO})_\text{pr} =
1/(2^{M-1} \sqrt{n_0}) = 2\sqrt{n_0}/N$, and with a Ramsey time that 
can exceed the laser noise limit $T_\text{max}$.




However, in general the estimation of the binary digits $Z_j$ is not perfect.
A rounding error occurs whenever $|\Phi_{j-1}^\text{est} - \Phi_{j-1}| > \pi/2$
(where $\Phi_j^\text{est}$ represents a suitable estimator derived from the $n_0$
measurement outcomes), leading to a wrong digit $Z_j$.
The variance contribution of such an event to the total measurement
uncertainty  is $(2\pi 2^{-j})^2$.
Since these errors are small and independent, we can approximate their total
variance contribution by the sum
 $ 	(\Delta\Phi_\text{LO})^2_\text{re} = P_\text{re}\sum_{j=1}^{M-1}
 	(2\pi 2^{-j})^2.
 $
The corresponding
probability in a single cycle is $P_\text{re} = 2\intop_{\pi/2}^{\infty}\d{\phi}
\rho(\phi)$,
where $\rho(\phi)$ is the Gaussian probability distribution of the error $\Phi_j^\text{est}
- \Phi_j$ with a width proportional to $1/\sqrt n_0$ \cite{SI}. 
Consequently, rounding errors can be
exponentially suppressed by choosing a sufficiently large value for $n_0$, and 
for $P_\text{re}\ll 1$, the total measurement uncertainty of this estimation
scheme is thus $(\Delta\Phi_\text{LO})^2 = (\Delta\Phi_\text{LO})_\text{pr}^2
+(\Delta\Phi_\text{LO})_\text{re}^2$.
In \cite{SI} we show that the optimal allocation of resources is achieved for the choice $n_{0}^{\text{opt}} \sim 
\frac{16}{\pi^2}\log\left(N\right)$ for which rounding errors are negligible, yielding the total measurement accuracy
 \bel
 \label{eq:M}
 	\Delta\Phi_\text{LO} \approx (\Delta\Phi_\text{LO})_\text{pr} = \frac{8}{\pi}\sqrt{\text{log}(N)}/N.
\eel
This measurement precision obtains the fundamental Cramer-Rao bound (up to a logarithmic correction resulting from the cost to suppress rounding errors) despite it being applicable to a general (typically large) phase.



So far we have assumed that $\Phi_\text{LO}\in[-\pi,\pi)$ in each cycle.
However, for realistic laser noise spectra there is always a finite probability that the LO phase $\Phi_\text{LO}$ lies outside the
interval $[-\pi,\pi)$ after the interrogation time. Such phase wraps of the laser phase itself add to the final measurement uncertainty in 
\refeq{eq:M} by the amount
$
	(\Delta\Phi_\text{LO})^2_\text{slip} =  (2\pi)^2P_\text{slip}.
$
The corresponding probability after a single Ramsey cycle is 
$P_\text{slip} = 2\intop_\pi^\infty \d{\phi} \rho_\text{LO}(\phi)$,
where $\rho_\text{LO}$ is the Gaussian prior distribution of $\Phi_\text{LO}$.
Its width depends directly on the product $ \gamma_\text{LO} T$, which puts a
constraint on the maximally allowed Ramsey time $T
\leq\frac{\pi^2}{4}\gamma_\text{LO}^{-1}[\log(\gamma_\text{LO}\tau N)]^{-1}$, and thus the achievable ADEV
$\sigma_y~(\propto 1/\sqrt T)$ 
 as we demonstrate in \cite{SI}.

As discussed above, this does not represent a fundamental limitation as we can extend the scheme by adding additional classical measurements with a shorter Ramsey periods (or alternatively, by employing dynamical decoupling techniques) \cite{Rosenband2013, Borregaard2013,ddc} to assess the number of phase slips of the laser phase itself $\hdots Z_{-3}Z_{-2}Z_{-1}Z_0$. As demonstrated in \cite{SI} this allows  extending the Ramsey time by a factor $k$ adding only a negligible number of atoms $N^{*}\approx \frac{8}{\pi^2} \log\left(k N^2\right)\log_2(k)\ll N$.
With all phase wraps counted correctly, the Ramsey time is only limited by
individual noise processes. The finite linewidth of the atomic clock transition
$\gamma_\text{ind}$ gives rise to the fundamental constraint
$T\leq\gamma_\text{ind}^{-1}/2^{M-1}$.
For averaging times smaller than this value,
$\tau\leq\gamma_\text{ind}^{-1}/2^{M-1}$, we can choose $T \approx \tau$, and
using the optimized value for $n_0$ found above the resulting clock stability is
obtained from \refeq{eq:Allan-variance}
\bel
	\label{eq:sigma(1)}
	\sigma_y(\tau)^{(1)} \approx	
	\frac{2}{\omega_0\tau}
	\frac{\sqrt{n_{0}^{\text{opt}}}}{N} \approx	\frac{8}{\pi\omega_0\tau}
	\frac{\sqrt{\text{log}(N)}}{N}.
\eel
It scales linearly with the averaging time $\tau$, and realizes the Heisenberg
bound of laser stability up to a logarithmic correction. In contrast, in the regime $\tau\geq\gamma_\text{ind}^{-1}/2^{M-1}$, $T$ is limited
by the presence of individual particle noise to a value $T\approx \gamma_\text{ind}^{-1}/2^{M-1}= 2 \gamma_\text{ind}^{-1}n_0/N$, and we find
\bel
	\label{eq:sigma(2)}
	\sigma_y(\tau)^{(2)} \approx
	\frac{1}{\omega_0}  \sqrt{\frac{\gamma_\text{ind}}{\tau N}}.
\eel
\refeq{eq:sigma(2)} represents the fundamental noise floor for laser stability
resulting from quantum metrological bounds in the presence of individual particle noise
\cite{Escher:2011fn}. As we have seen, the proposed protocol reaches this optimal value rapidly
after the averaging time $\tau_0 \sim   \gamma_\text{ind}^{-1} \text{log}(N)/N$ (cf. \reffig{fig:sigma_tau}), $N/\text{log}(N)$ times faster than any classical scheme.

In the following, we benchmark the stability of our protocol against different approaches by
comparing the lowest achievable ADEV as a function of averaging
time $\tau$ (cf. \reffig{fig:sigma_tau}). 
First, we consider the standard procedure in which all atoms are interrogated
in an uncorrelated fashion. The scheme is identical to $N$
independent measurements of $\Phi_\text{LO}$, and therefore the  ADEV
is limited by the standard quantum limit: $\sigma_y \sim \frac{1}{\omega_0
\tau\sqrt{N}}$ for $\tau < \gamma_\text{LO}^{-1}$.
Since the Ramsey time is limited by the LO noise to $T<\gamma_\text{LO}^{-1}$ due to uncorrected phase wraps, this naive protocol fails to achieve the fundamental bound \refeq{eq:sigma(2)} in the long time limit
$ \tau > \gamma_\text{LO}^{-1}$, and we find the suboptimal ADEV
  $\sigma_y(\tau) \sim
\frac{1}{\omega_0}\sqrt{\frac{\gamma_\text{LO}}{\tau N}}$. 
Second, we discuss the recently published classical protocol which
interrogates the LO with uncorrelated atoms for exponentially increasing Ramsey
times in each cycle \cite{Borregaard2013, Rosenband2013}. This protocol can be
understood as the classical part of the cascaded interrogation proposed here
($j\leq0$).
It eliminates the constraint of the LO linewidth, and allows to extend the interrogation time $T$ to its maximum value, enabling a linear scaling with $\tau$ up to the point where the fundamental bound
(\ref{eq:sigma(2)}) is reached.
However, using an uncorrelated interrogation, the scheme displays a standard-quantum-limited scaling (i.e.
$\propto1/\sqrt{N}$), for short averaging times. 
 
The above analysis illustrates the quantum gain of the proposed clock
operation protocol using cascaded GHZ states. As compared to the best known
classical scheme, our scheme provides a $\sqrt{N/\text{log}(N)}$ enhancement for
short averaging times. As a result it reaches the fundamental noise floor for
laser stability in the presence of single particle decoherence
[\refeq{eq:sigma(2)}] $\sim N/{\text{log}(N)}$ times faster. In the limit of
$N\rightarrow \infty $, our scheme attains this optimal stability allowed by
quantum mechanics for all values of $\tau$.
This results identifies the possible advantage of using entanglement
previously debated in the literature
\cite{Wineland1998,Huelga1997,LouchetChauvet:2010fs,Rosenband2012_numerical,Borregaard2013_nearHeisenberg,Andre2004,Buzek1999,Meiser:2008fo}:
While the long term limitation may be set by atomic decoherence, entangled atoms
reaches this limit faster thus improving the bandwidth of the stable oscillator.
Our results motivate the development of atomic clocks based on entangled ions
and neutral atoms. Furthermore this motivates and lays the foundations for a
network of quantum clocks which operates by interrogating entangled states of all atoms in the network
collectively, therefore achieving a stability set by the Heisenberg limit of all
atoms \cite{tbp}. 
Such a network can be used to construct a real-time world clock. Modifications
of the scheme, such as employing optimized numbers of copies on each level of the cascade, or conditional rotations of the measurement basis,
might allow to overcome the logarithmic correction in the achievable stability,
and are subject to future investigations.

We are grateful to Till Rosenband and Johannes Borregaard for enlightening
discussions. This work was supported by NSF, CUA, ITAMP, HQOC,
JILA PFC, NIST, DARPA QUSAR, the Alfred P. Sloan
Foundation, the Quiness programs, ARO MURI, and
the ERC grant QIOS (grant no. 306576); MB acknowledges support from NDSEG and NSF GRFP.

\appendix 
\section*{SUPPLEMENTARY INFORMATION}

\section{Figure of merit: Allan-variance}
Provided $N$ qubits, we aim to devise an efficient interrogation scheme that
provides input for the feedback mechanism, using Ramsey spectroscopy. After the $k$th
Ramsey cycle, of length $T$, an estimate $\Phi^\text{est}_\text{LO}(t_k)$ is
obtained for the accumulated phase of the LO, $\Phi_\text{LO}(t_k) =
\intop_{t_k-T}^{t_k}\d{t}\delta\omega(t)$, ($t_k = kT$, $k=1,2\dots$, and $\delta\omega(t) =\omega(t)- \omega_0$), which differs
from the real value by $\Delta\Phi_\text{LO}(t_k) =
\Phi^\text{est}_\text{LO}(t_k) - \Phi^\text{real}_\text{LO}(t_k)$. 
Using the obtained estimate, the feedback
mechanism corrects the phase or frequency of the LO after every cycle, and
thus creates a LO signal with better stability.
The figure of merit for stability is the Allan-variance,
$
	\sigma_y^2(\tau) = \frac{1}{\omega_0^2}
	\ev{\delta\bar\omega^2(t_0)}_{t_0}
$
where $\delta\bar\omega(t_0)$ denotes the time-average of $\delta\omega(t_0 +
t)$ over $t\in[0,\tau]$, where $\tau$ is the available averaging time,
$\ev{\;}_{t_0}$ stands for time-average over $t_0$, which is much longer than
 $\tau$, $\omega_0$ is the frequency of the chosen
clock transition. Consequently, one readily shows that the Allan variance can be written as,
\bel
	\sigma_y^2(\tau) = \frac{1}{\omega_0^2\tau^2}
	\sum_{i=1}^{\tau/T}\sum_{j=1}^{\tau/T} \Big\langle\Delta\Phi_\text{LO}(t_0 +
	iT)\Delta\Phi_\text{LO}(t_0 + jT)\Big\rangle_{t_0}.
\eel
By assuming that $\Delta\Phi_\text{LO}$ is a stationary random process, we
substitute the average over $t_0$ with the average over many realizations. With the notation
$\Delta\Phi_\text{LO}(t_0 + jT) = \Delta\Phi_{\text{LO},j}$, we can write this
average as
\bel
	\label{eq:DeltaPhi}
	\ev{\Delta\Phi_{\text{LO},i}\Delta\Phi_{\text{LO},j}} \approx
	\ev{\Delta\Phi^2_\text{LO}} \delta_{ij},
\eel
where we further used a white noise assumption, such that phase accumulations in consecutive Ramsey cycle are approximately
uncorrelated. Also for realistic $1/f$ laser noise spectra, numerical studies show that this factorization assumption leads to only negligible corrections. Earlier
results show that this is the case for initial LO frequency noise spectra,
$S_\nu(f)$ that are less divergent than $1/f^2$ at low frequencies
\cite{Andre2005}. As a result, the Allan-variance simplifies to,
\bel
\label{eq:Allan-var}
	\sigma_y^2(\tau) = \frac{1}{\omega_0^2\tau T} \ev{\Delta\Phi^2_\text{LO}},
\eel
linking the achieving stability directly to the frequency measurement uncertainty during the interrogation.
\refeq{eq:Allan-var} serves as our starting point in finding the optimal
measurement protocol that minimizes $\sigma_y^2(\tau)$ for fixed $N$ and $\tau$. In the following, we investigate and compare different classical and quantum mechanical strategies for the interrogation of the (from cycle to cycle fluctuating) quantity $\Phi_\text{LO}$, and we demonstrate that the proposed interrogation protocol using cascaded GHZ states is optimal up to a logarithmic correction.

\section{Single-step Uncorrelated ensemble}
\label{sec:P1}
First, we consider the case of naive interrogation using a Ramsey protocol with $N$
 uncorrelated  atoms. 

\subsection{Sub-ensembles and projection noise}
Single ensemble Ramsey spectroscopy is limited to estimating
either the real or the imaginary part of $e^{i\Phi_\text{LO}}$. However, by
dividing the available qubits into two sub-ensembles, $X$ and $Y$, preparing
their individual qubits in different states,
\bal
	\label{eq:X_ensemble_uncorr}
	X &:\quad& [\ket{0} + \ket{1}]/\sqrt{2},\\
	\label{eq:Y_ensemble_uncorr}
	Y &:\quad& [\ket{0} + i\ket{1}]/\sqrt{2},
\eal
and performing the same Ramsey measurement on them, we can get
 estimates on both the real and imaginary parts of $e^{i\Phi_\text{LO}}$ and
deduce the value of $\Phi_\text{LO}$ up to $2\pi$ shifts, instead of $\pi$.
At the end of the free evolution time, each qubit in ensemble $X$ ($Y$) is
measured in the $x$-basis ($\ket{\pm} = [\ket{0}\pm\ket{1}]/\sqrt{2}$) and yields the
'+' outcome with probability $p_x = [1-\cos\Phi_\text{LO}]/2$ ($p_y =
[1-\sin\Phi_\text{LO}]/2$).
%

After performing the measurement with $N$ total qubits, we obtain
$\Phi_\text{LO}^\text{est}$ from the estimates of $p_x$ and $p_y$. Since both
provide information on
$\Phi_\text{LO}^\text{est}$ equivalent of $N/2$ measurement bits, this results
in a total information of $N$ measurement bits, which gives an uncertainty of
\bel
	\label{eq:Projection}
	\ev{\Delta\Phi^2_\text{LO}}_\text{pr} = \frac{1}{N},
\eel
up to $2\pi$ phase shifts, that are fundamentally undetectable.
This method is identical to the one described in \cite{Rosenband2013}.

\subsection{Effects of laser fluctuations: Phase slip errors}
Random fluctuations in the laser frequency (characterized by the laser spectrum noise spectrum $S_\nu(f) = {2\gamma_\text{LO}}/{f}$) result in the fact that the laser phase itself has to be considered as a random variable after each cycle. 
 Whenever in a given cycle the phase $\Phi_\text{LO}(t_k)$ falls outside the
 interval $[-\pi,\pi]$, the aforementioned technique leads to an estimate deviating from the true value by $\sim2\pi$. As the variance $s^2$ of the prior distribution of  $\Phi_\text{LO}$ grows with the interrogation time $T$ (one finds $s^2= \gamma_\text{LO}T$ ($s^2= (\gamma_\text{LO}T)^2$) for a white ($1/f$) noise frequency spectrum, where $\gamma_\text{LO}$ denotes the laser linewidth of the free-running
(non-stabilized) LO) these undetected \textit{phase slips} pose a fundamental limitation on the allowed Ramsey time $T$, and thus on the overall achievable laser stability.

If we assume a constant rate of phase diffusion, resulting in a Gaussian
prior distribution of  $\Phi_\text{LO}$,
the probability of a phase slip single cycle of length
$T$ can be estimated as
\bal
	\label{eq:Pslip in T}
	\PP_\text{slip} &=&  2\intop_\pi
	^{\infty}\d{\Phi_\text{LO}^\text{real}}
	\frac{1}{\sqrt{2\pi s^2}}
	\exp\left[-\frac{(\Phi_\text{LO}^\text{real})^2}{2s^2}\right] \\\nonumber
	&=& 1-\text{erf}\left(\frac{\pi}{\sqrt 2 s}\right)
	\\
	&=&
	\label{eq:Pslip in T2}
	\left(\frac{\sqrt{2}s}{\pi^{3/2}} +
	\mathcal{O}(s^2)\right)\exp\left[-\frac{\pi^2}{2s^2}\right],
\eal
where erf denotes the error function.
 As a phase slip in an early Ramsey cycle will remain undetected in the following
cycles, its error contribution will accumulate over the total averaging time
$\tau$, in the worst case by a factor $\tau/T$.
Using this upper bound, and assuming $\PP_\text{slip  }\ll1$ we write the variance contribution of phase slips as
\begin{align}
	\label{eq:Phase slip}
	\ev{\Delta\Phi^2_\text{LO}}_\text{slip} =& (2\pi)^2 \frac{\tau}{T} \PP_\text{slip  } \\\nonumber
	\approx&(2\pi)^2\frac{\tau}{T}\frac{\sqrt{2}}{\pi^{3/2}}\sqrt{\gamma_\text{LO} T}
	\exp\left[-\frac{\pi^2}{2\gamma_\text{LO}T}\right],
\end{align}
where the $(2\pi)^2$ prefactor sets the absolute contribution of a manifested
slip event to $\pm 2\pi$, and in the second step we approximated $\PP_\text{slip}$ with the
first term of its asymptotic series from \refeq{eq:Pslip in T2}.

\subsection{Optimal Ramsey time}
While \eqref{eq:Allan-var} suggests increasing Ramsey times improve the laser stability, we have seen in the previous section that they also lead to an increased occurrence of phase slips yielding a significant contribution to the measurement uncertainty.

In order to find the optimal Ramsey time
we add the contributions from quantum projection noise
[\refeq{eq:Projection}] and phase slip noise [\refeq{eq:Phase slip}] under the assumption that the
probability of phase slips is small, and obtain an expression for the
Allan-variance, 
\bel
\label{eq:EK9}
\sigma_y^2(\tau) = \frac{1}{\omega_0^2 \tau}\Gamma,
\eel
 where
\bel
\label{eq:abc}
	\Gamma = \frac{1}{TN} +
	\sqrt{32\pi}
	\frac{\tau\gamma_\text{LO}^{1/2}}{T^{3/2}}
	\exp\left[-\frac{\pi^2}{2\gamma_\text{LO}T}\right].
	\eel
In order to find the optimal Ramsey time $T_\text{opt}$, that minimizes
$\Gamma$, we introduce the new variable $x = \frac{2}{\pi^2}\gamma_\text{LO}T$,
and write
\bel
	\Gamma = \frac{2}{\pi^2}\frac{\gamma_\text{LO}}{Nx} + \frac{16}{\pi^{5/2}}
	\frac{\tau\gamma_\text{LO}^2}{x^{3/2}} e^{-1/x}.
\eel
Taking the derivative with respect to $x$ results in
\bel
	\frac{d}{dx}\Gamma = -\frac{2}{\pi^2}\frac{\gamma_\text{LO}}{Nx^2} +
	\frac{16}{\pi^{5/2}}
	\tau\gamma_\text{LO}^2\left(-\frac{3}{2}\frac{1}{x^{3/2}} +
	\frac{1}{x^{7/2}}\right) e^{-1/x},
\eel
which, after using the (self-consistent) assumption $x_\text{opt} \ll 1$, results in the following
transcendental equation for $x_\text{opt}$,
\bel
	x_\text{opt}^{3/2} = ((8/\sqrt{\pi}) \gamma_\text{LO}\tau N)
	e^{-1/x_\text{opt}}.
\eel 
Below,
 we provide the derivation of the asymptotic solution,
\bal
	x_\text{opt} &=& [\log((8/\sqrt{\pi}) \gamma_\text{LO}\tau N)]^{-1} \approx
	[\log(\gamma_\text{LO}\tau N)]^{-1},
\eal
yielding directly
\bal
\label{eq:T_opt}
	T_\text{opt} &\approx& \gamma_\text{LO}^{-1}\frac{\pi^2}{2}
	[\log(\gamma_\text{LO}\tau N)]^{-1}.
\eal
Self-consistently we confirm that already for values $\gamma_\text{LO}\tau N\geq 10^4$, the approximation 
in \refeq{eq:Phase slip} is well justified, so that the above value represents a true local minimum. 
For larger values of $T$ the phase slip errors grow rapidly, and numerical studies confirm that \refeq{eq:T_opt} indeed represents a global minimum.

The optimal interrogation time is mainly set by the LO coherence time
$\gamma_\text{LO}^{-1}$, and shows a weak dependence on the total number of qubits
$N$, and the averaging time $\tau$ (Note, that
if we model the LO with a $1/f$ frequency noise spectrum, only the
exponent of the $\log$ term changes to $-1/2$ in this result). Using this optimized Ramsey time we find for the 
minimal value of $\Gamma$ is
\bal
	\Gamma_\text{min} &=&
	\frac{2}{\pi^2}\frac{\gamma_\text{LO}}{Nx_\text{opt}}
	+
	\frac{16}{\pi^{5/2}}
	\frac{\tau\gamma_\text{LO}^2}{x_\text{opt}^{3/2}} e^{-1/x_\text{opt}} 	\\
	&=&
	\frac{2}{\pi^2}\frac{\gamma_\text{LO}}{N}
	\left(\frac{1}{x_\text{opt}} + 1\right) 
	\\
	&\approx& 
	\label{eq:gamma_eff}
	\frac{2}{\pi^2}
	\frac{\gamma_\text{LO}\log(\gamma_\text{LO}\tau N)}{N}.
\eal

This result is valid as long as the averaging time $\tau$ is longer than the
proposed $T_\text{opt}$ from \refeq{eq:T_opt}. If this is not the case, then
$T_\text{opt} = \tau$, the phase slip noise becomes
negligible, and we end up with
\bel
	\label{eq:gamma_eff_short_tau}
	\Gamma_\text{min} = 
	\frac{1}{\tau N}.
\eel

We approximate the
crossover region (around $\tau\sim \gamma_\text{LO}^{-1}$) by adding leading
terms from \refeq{eq:gamma_eff} \& (\ref{eq:gamma_eff_short_tau}) and obtain
\bel
	\label{eq:sigma_y^2 boxed}
	[\sigma_y(\tau)]_\text{min}
	\approx \frac{1}{\omega_0 \sqrt{N\tau}} \sqrt{\frac{1}{\tau}
	+ \frac{2}{\pi^2}\gamma_\text{LO}\log(\gamma_\text{LO}\tau
	N)}.
\eel

In summary, in the region $\tau<T_\text{opt}$, the LO noise is negligible leading to a linear scaling of the ADEV with the total averaging time $\tau$. For large averaging times $\tau>T_\text{opt}$, phase slips of the laser phase pose a limitation to the maximal possible Ramsey time which results in a $1/\sqrt \tau$ scaling of the laser stability. Since we employ uncorrelated atoms, the ADEV displays in both regimes the $1/\sqrt{N}$ scaling of the standard quantum limit (SQL).
As modern atomic clocks typically are laser noise limited, $\gamma_\text{LO}\gg \gamma_\text{ind}$ (where $\gamma_\text{ind}$ represents the clock atom linewidth), we neglected the effects of individual atomic dephasing in the above considerations.

\section{Cascaded interrogation using GHZ states}
In this Section, we discuss the possibility of using quantum correlated
states, namely GHZ states of the form
\bel
	[\ket{\mathbf{0}} + e^{i\chi} \ket{\mathbf{1}}]/\sqrt{2},
\eel
where $\ket{\mathbf{0}}$ and $\ket{\mathbf{1}}$ are product states of all qubits
being in $\ket{0}$ or $\ket{1}$, respectively, and $\chi$ will be referred to as
the phase of the GHZ state. Such a state, once prepared, is more sensitive to
the accumulated phase of the LO, $\Phi_\text{LO}$, by a factor of $N'$, the
number of qubits entangled:
\bel
	\left(\prod_{j=1}^N \hat U_j\right) \left[\ket{\mathbf{0}} + e^{i\chi}
	\ket{\mathbf{1}}\right]/\sqrt{2} 
	= [\ket{\mathbf{0}} + e^{i(\chi + N'\Phi_\text{LO})}
	\ket{\mathbf{1}}]/\sqrt{2},
\eel
where $\hat U_j = \ket{0}\bra{0} + e^{i\Phi_\text{LO}}\ket{1}\bra{1}$ is the
time propagation operator for the interrogation time acting on the $j$th qubit. This property promises an
enhancement in phase resolution, and therefore a better stability for quantum clocks.

\subsection{Parity measurement}
Using the idea with the two
sub-ensembles [see \refeq{eq:X_ensemble_uncorr} and
(\ref{eq:Y_ensemble_uncorr})], we imagine dividing the qubits into two equal
groups, and preparing two separate GHZ states:
\bal
	\ket{X} &:=& [\ket{\mathbf{0}} + \ket{\mathbf{1}}]/\sqrt{2},\\
	\ket{Y} &:=& [\ket{\mathbf{0}} + i\ket{\mathbf{1}}]/\sqrt{2},
\eal
each entangling $N'$ qubits.
After the free evolution time, we imagine measuring each qubits in
the $x$-basis ($\ket{\pm} = [\ket{0} \mp \ket{1}]/\sqrt{2}$) separately. In this
basis, the above states are written as
\bel
	\left[\left(\frac{\ket{+} - \ket{-}}{\sqrt{2}}\right)^{\otimes N'}
	+  e^{i\phi_\nu}\left(\frac{\ket{+} +\ket{-}}{\sqrt{2}}\right)^{\otimes
	N'}\right]/\sqrt{2},
\eel
where $\phi_\nu = \chi_\nu + N'\Phi_\text{LO}$, for $\nu\in\{x,y\}$ and
$\chi_x = 0$, while $\chi_y = \pi/2$ for the two groups, respectively. The above
state can be written as
\bel
	\frac{1}{2^{(N'+1)/2}}\sum_{\mathbf{q}\in\{+,-\}^{\times N'}}
	\left[ \left(\prod_{j=1}^{N'} q_j \right) + e^{i\phi_\nu}\right]
	\ket{q_1,q_2,\dots q_{N'}}.
\eel
Once
the qubits are measured one by one, the probability to measure a certain outcome
$\mathbf{q} = (q_1, q_2, \dots q_{N'})$, ($q_j \in \{+,-\}$) is
\bel
	\PP(\mathbf{q}) = \frac{1}{2^{N' +1}} |1+ p(\mathbf{q})e^{i\phi_\nu}|^2,
\eel
where $p(\mathbf{q}) = \prod_{j=1}^{N'} q_j$ is the parity of the sum of all
measurement bits. This parity is the observable that is sensitive to the
accumulated phase, since its distribution is
\bel
	\PP(p=\pm 1) = \frac{1\pm \cos(\phi_\nu)}{2}.
\eel
This is identical to the parity measurement scheme described in
\cite{Bollinger1996}.
By interrogating $n_0/2$ instances of $\ket{X}$ and $\ket{Y}$, respectively, we
can measure the phase of the GHZ state, $N' \Phi_\text{LO}$, with uncertainty
$1/\sqrt{n_0}$, since each instance provides a single measurement bit, which can
be combined the same way as we described in the case of uncorrelated ensembles.
The resulting measurement uncertainty, $\Delta\Phi_\text{LO}$, is
\bel
	\label{eq:projection_GHZ_1}
	\ev{\Delta\Phi^2_\text{LO}}_\text{pr} = \frac{1}{(N')^2 n_0} =
	\frac{n_0}{N^2},
\eel
which is a factor of $N/n_0$ smaller than the
variance contribution of projection noise for the uncorrelated ensemble
protocol, ($N = n_0 N'$).

\subsection{Failure of the maximally entangled GHZ}
Motivated by the increased phase resolution provided by the interrogation of GHZ
states, we evaluate the stability of such a protocol. We find that it fails to
provide improvement compared to the single-step uncorrelated protocol due to an
increased phase slip rate. This is in agreement with earlier results
\cite{Wineland1998,Rosenband2012_numerical}.

The probability, that the phase accumulated by $\ket{X}$ ($\ket{Y}$) during the
interrogation time $T$, $N'\Phi_\text{LO}$ lies outside the interval
$[-\pi,\pi]$, is
\bel
	\PP_\text{slip} = 2\intop_{\pi/N'}^\infty \d{\Phi_\text{LO}^\text{real}}
	\frac{1}{\sqrt{2\pi \gamma_\text{LO}T}}
	\exp\left[-\frac{(\Phi_\text{LO}^\text{real})^2}{2\gamma_\text{LO}T}\right],
\eel 
which, due to the much lower slipping threshold of $\pi/N'$ [instead of $\pi$
in the uncorrelated case, compare \refeq{eq:Pslip in T}] will become significant
for much shorter $T$ cycle times.
The resulting variance contribution (following the same argument as before) is
\bal
	&&\ev{\Delta\Phi^2_\text{LO}}_\text{slip} =\nonumber\\
	&& 
	\label{eq:slip_GHZ_1}
	\qquad = \sqrt{32\pi}\frac{\tau}{T}
	\sqrt{\gamma_\text{LO}T}N'
	\exp\left[-\frac{\pi^2 }{2\gamma_\text{LO}T(N')^2}\right].
\eal
Neglecting the individual qubit noise by the same argument as before, we simply add the contributions
from \refeq{eq:projection_GHZ_1} and \refeq{eq:slip_GHZ_1} to obtain the
Allan-variance, $\sigma_y^2(\tau) = \frac{1}{\omega_0^2\tau} \Gamma$, where
\bel
	\Gamma =  \frac{1}{NN' T} +
	\sqrt{32\pi} \frac{\tau \gamma_\text{LO}^{1/2}}{T^{3/2}}N' 
	\exp\left[-\frac{\pi^2 }{2\gamma_\text{LO}T(N')^2}\right].
\eel
After 
optimizing $T$, we find
\bel
	T_\text{opt} \approx \frac{\pi^2}{2}\frac{1}{\gamma_\text{LO} (N')^2}
	\frac{1}{\log[\gamma_\text{LO}\tau N (N')^3]},
\eel
which results in the minimal Allan-variance,
\bel
	[\sigma_y(\tau)]_\text{min} \approx
	\frac{1}{\omega_0}\frac{\sqrt{2}}{\pi}\sqrt{\frac{\gamma_\text{LO}N' 
	\log[\gamma_\text{LO}\tau N (N')^3]}{\tau N}},
\eel
which is at least a factor of $\sqrt{N'}$ \emph{bigger} than the smallest obtainable
Allan-variance with the single-step uncorrelated protocol [\refeq{eq:gamma_eff}].
In case of a $1/f$ LO frequency noise spectrum, $T_\text{opt} \propto
\frac{1}{N}$ (up to logarithmic terms), and the resulting Allan-variance is
equal to \refeq{eq:gamma_eff}, up to logarithmic corrections, yielding
effectively no advantage over the uncorrelated scheme.

\subsection{Cascaded GHZ scheme}
\label{sec:casGHZ}
As demonstrated in the previous Section, single GHZ states fail to improve clock
stability because the increase in sensitivity to the laser detuning, at the
same time, leads to a
drastic increase of phase slip errors originating from laser frequency fluctuations. 
These fluctuations, however, affect all clock qubits in identical manner, and
therefore represents a \textit{collective noise}. As such (and unlike, e.g., the individual dephasing of the clock qubits),
they do not represent a fundamental limitation for the
phase estimation.
In the following, we show that this
problem can be efficiently addressed using a cascade of GHZ states of varying
sizes (and classical states with varying interrogation times) in an incoherent version of the phase estimation algorithm
\cite{Nielsen_Chuang}. 
To this end, we
reformulate the problem of estimating $\Phi_\text{LO}$ in a more suitable
language.

The laser phase after a given Ramsey cycle can be expressed in a base-$D$
numeral system as
\bel
	\label{eq:D-digits}
	(\Phi_\text{LO} + \pi)/2\pi= \sum^\infty_{j=-\infty} Z_j / D^{j},
\eel
with base-$D$ digits $Z_j \in\{0,1, \hdots, D-1\}$.
Let us for the moment assume that the laser phase $\Phi_\text{LO}\in
[-\pi,\pi]$, such that $(\Phi_\text{LO} + \pi)/2\pi= \sum^\infty_{j=1} Z_j /
D^{j}\equiv 0.Z_1Z_2Z_3\hdots$.

Provided with $N$ qubits, we imagine dividing them into $M$
different groups, the $j$th group ($j=0,1,\dots M-1$) contains $n_0$
instances of GHZ states with $D^j$ number of entangled qubits.
One readily shows that a GHZ state consisting of $D^{M-1}$ particles picks up
the phase
\begin{align}
	\Phi_{M-1}&= D^{M-1} \Phi_\text{LO} \mod [-\pi,\pi] \\
	 &= 2\pi (0.Z_{M}Z_{M+1}Z_{M+2}\hdots)-\pi,
\end{align}
which depend only on digits $Z_{M}$ and higher of the laser phase to be
measured.
This insensitivity of the GHZ state with regard to the lower digits $Z_1$ to
$Z_{M-1}$ restates the problems of phase slips. Only if the latter are known, a
measurement of the phase of the GHZ state $\Phi_{M-1}$ yields useful information
to determine $\Phi_\text{LO}$. In other words, the natural number $Z_1Z_2\hdots
Z_{M-1}$ represents the number of phase slips of the largest group of GHZ states
($j=M-1$).
These lower digits can be determined one by one from the accumulated phases
$\Phi_j = D^{j} \Phi_\text{LO} \mod [-\pi,\pi]$ of the smaller GHZ ensembles $j=0, \hdots, M-2$ by using the relation
\begin{align}
	\label{eq:Z_j_measuring}
	 [D&(\Phi_{j-1} + \pi) - (\Phi_j +
	\pi)]/(2\pi)\\\nonumber
	&=(Z_j.Z_{j+1}Z_{j+2}\hdots) - (0.Z_{j+1}Z_{j+2}\hdots) = Z_j.
\end{align}

Combining all measurement results we find that the best estimate for
$\Phi_\text{LO}$ is given by
\bel
\Phi_\text{LO}^\text{est} = 2\pi\sum_{j=1}^{M-1} Z^\text{est}_j /D^{j} +
\Phi^\text{est}_{M-1}/D^{M-1},
\eel
the precision of which is mostly determined by the uncertainty of the phase of
the last group ($j=M-1$), which contains the GHZ states with the most entangled
qubits. Since there are $n_0$ independent instances of these GHZ states, their
phase is known up to the uncertainty,
$
	\ev{\Delta\Phi^2_{M-1}}_\text{pr} = \frac{1}{n_0} \approx \frac{\delta
	D^{M-1}}{N }
$
, where $\delta = \frac{D}{D-1}$, and therefore we find
\bel
	\label{eq:projection_GHZ 2}
	\ev{\Delta\Phi^2_\text{LO}}_\text{pr} =
	\frac{\ev{\Delta\Phi^2_{M-1}}_\text{pr}} {D^{2(M-1)}}  =
	\frac{n_0\delta^2}{N^2}.
\eel
This would be the total uncertainty if we could tell with certainty that all
phase slips of the lower levels had been detected correctly. However, the occurrence of an error in the estimation of any $Z_j$ (in the following referred to as \textit{rounding error}) has non-zero probability. 

\subsection{Rounding errors: finding the optimal $n_0$}
\label{sec:RE}
If $\Phi_j$ is determined
with poor precision, the estimate of $Z_{j+1}$ will have a significant uncertainty,
causing the final estimate of $\Phi_\text{LO}$ to be uncertain as well. Whenever
$|\Phi_j^\text{est} - \Phi_j^\text{real}| > \pi/D$, we make a mistake by under-
or overestimating the digit $Z_{j+1}$. To minimize the effect of
this error, we need to optimize how the qubits are distributed on  various
levels of the cascade. In other words, for a given total particle number $N$ and basis $D$
we need to find the optimal value of $n_0$, the number of copies of GHZ states in each step 
\footnote{In principle, the clock stability can further be improved by employing different numbers of copies in each step of the Cascade. However, this possibility will not be pursuit in this work.}.

The probability that a rounding error occurs during the estimation of
$Z_{j}$ is
\bal
	\PP_\text{re} &=& 2\intop_{\pi/D}^\infty \d{\phi}
	\rho(\phi - \Phi_j^\text{real}) \leq 2\intop_{\pi/D}^\infty \d{\phi} n_0^{3/2}
	\exp\left[-\frac{n_0\phi^2}{2}\right] \nonumber\\
	&\approx  &
	\frac{2}{\pi} n_0^{1/2} D \exp\left[-\frac{n_0\pi^2}{2D^2}\right]
	,
\eal
where $\phi = \Phi_{j-1}^{\text{est}} - \Phi_{j-1}^\text{real}$, and $\rho$ is
the conditional distribution of $\Phi_{j-1}^\text{est}$ for given
$\Phi_{j-1}^\text{real}$.
The employed upper bound is obtained in 
the last section, 
with the
assumption $\gamma_\text{LO}/\gamma_\text{ind} \gg N/n_0$ ($\gamma_\text{ind}$
is the individual qubit dephasing rate), so that the projection noise is the
dominant noise term.
Due to the fixed value of $n_0$ accross different levels of the cascade, this
probability is independent of $j$, however the phase shift imposed on
$\Phi_\text{LO}$ by a manifested rounding error of $Z_{j}$ is $2\pi D^{-j}$ ($j=1,\dots M-1$), as rounding errors early in the cascade are more harmful than later ones.
This results in the total variance contribution,
\bal
	\label{eq:Rounding_GHZ 2}
	&&\ev{\Delta\Phi^2_\text{LO}}_\text{re}  =
	\PP_\text{re} \sum_{j=1}^{M-1}(2\pi
	D^{-j})^2\\
	&&\approx 8\pi
	\sqrt{\frac{N}{\delta}} D^{-\frac{M-3}{2}}
	\exp\left[-\frac{n_0\pi^2}{2D^2}\right]
	\frac{1}{D^2-1}.
\eal

By adding the two error contributions from \refeq{eq:projection_GHZ 2}
and \refeq{eq:Rounding_GHZ 2},
we obtain the total uncertainty, $\ev{\Delta\Phi^2_\text{LO}}$ and the
corresponding Allan-variance [according to \refeq{eq:Allan-var}]
\bal
	\sigma_y^2(\tau) &=& \frac{1}{\omega_0^2\tau}\left[\frac{\delta}{N T
	D^{M-1}}
	\right.
	\nonumber\\
	&&
	+
	 \left.\frac{8\pi}{T}
	\sqrt{\frac{N}{\delta}} D^{-\frac{M-3}{2}}
	\exp\left[-\frac{n_0\pi^2}{2D^2}\right]\frac{1}{D^2-1}
	\right]\nonumber
	\\
	\label{eq:4 Gamma}
	&=:& \frac{1}{\omega_0^2\tau}\left[\Gamma_1 + \Gamma_2\right]
\eal

We find the optimal value of $n_0$ by minimizing this quantity. Introducing the new variable $x \equiv \frac{2D^2}{n_0\pi^2}$, and using $ n_0 \approx N/(\delta D^{M-1})$ we write
\bel
	\Gamma_1 + \Gamma_2 = \frac{2}{\pi^2}\frac{\delta^2 D^2}{N^2 T  x} +
	\frac{\sqrt{128} D^2}{T(D^2-1)} \frac{1}{x^{1/2}}
	\exp\left[-\frac{1}{x}\right]
\eel
Taking the derivative with respect to $x$ and equating it with 0, while using
the (self-consistent) assumption $x_\text{opt} \ll 1$, results in the condition $\Gamma_2 \approx
x_\text{opt} \Gamma_1  \ll \Gamma_1$ and the transcendental equation
\bel
	x_\text{opt}^{1/2} \approx \frac{\sqrt{32} \pi^2 N^2}{\delta^2
	(D^2-1)} \exp\left[-\frac{1}{x_\text{opt}}\right]
\eel
for $x_\text{opt}$.
The asymptotic solution in the case of $x_\text{opt} \ll 1$ is 
\bal
	x_\text{opt} &\approx& \left[\log\left(\frac{\sqrt{32} \pi^2
	N^2}{\delta^2 
	(D^2-1)}\right)\right]^{-1}\sim\left[\log\left(N^2\right)\right]^{-1},
\eal
yielding directly the optimal number of instances of GHZ states per level
\bal
\label{eq:EK6}
	n_0^\text{opt} \sim \frac{2}{\pi^2} D^2 \log\left(N^2\right).
\eal
This choice guarantees rounding errors yield a negligible contribution to the total measurement uncertainty, and we find
for the corresponding
value of $\Gamma_1 + \Gamma_2$ 
\begin{align}
	\label{eq:Gamma1_Gamma2}
	[\Gamma_1 + \Gamma_2]_\text{min} \approx& \Gamma_1(x_\text{opt}) \\\nonumber
	=&\frac{n_0^\text{opt}\delta^2}{N^2T}
	\sim
	\frac{2}{\pi^2}\frac{\delta^2 D^2}{N^2 T} 
	\log\left(N^2\right),
\end{align}
where the factor $\delta=D/(D-1) \in (1,2]$.
Obviously, the use of a binary basis  ($D=2$) is optimal, and the effect of rounding errors lead to a logarithmic correction to the Heisenberg limit.

\subsection{Phase slip errors: limitations to the Ramsey time $T$ from laser noise}
\label{sec:PSE}
Although the cascade is designed to detect phase slips at levels $j=1,2\dots
M-1$, when we relax the condition $\Phi_\text{LO}\in [-\pi,\pi]$, and allow
$\Phi_\text{LO} \in (-\infty, + \infty)$, the possible phase slips of level
$j=0$ ($Z_0$) remains undetected.
Once this happens, it introduces a $2\pi$ phase shift in $\Phi_\text{LO}$, and
therefore contributes to its overall uncertainty with
\bal
	\ev{\Delta\Phi^2_\text{LO}}_\text{slip} &=& (2\pi)^2\frac{\tau}{T}
	\PP_\text{slip} = \nonumber
	\\
	\label{eq:Slip_GHZ 2}
	&=& 
	\sqrt{32\pi} \frac{\tau \gamma_\text{LO}^{1/2}}{T^{1/2}}
	\exp\left[-\frac{\pi^2}{2\gamma_\text{LO}T}\right],
\eal
where we
assumed $\gamma_\text{LO}T \ll 1$.
This adds an extra noise term $\Gamma_3 := 
\ev{\Delta\Phi_\text{LO}^2}_\text{slip}/T$ to the already optimized $[\Gamma_1 +
\Gamma_2]_\text{min}$ expression, yielding
\bal
	&& [\Gamma_1 + \Gamma_2]_\text{min} + \Gamma_3 = \nonumber
	\\
	&& \quad =
	\frac{2}{\pi^2}\frac{\delta^2n_0^\text{opt}}{N^2y} +  \frac{16}{\pi^{5/2}}\tau
	\gamma_\text{LO}^2 \frac{1}{y^{3/2}}\exp\left[-\frac{1}{y}\right],\quad\quad
\eal
where $y = \frac{2}{\pi^2}\gamma_\text{LO}T$. After taking the derivative
with respect to $y$ and equating it with zero, the assumption $y_\text{opt}\ll
1$ results in the condition $\Gamma_3 \approx y_\text{opt} [\Gamma_1 +
\Gamma_2]_\text{min} \ll [\Gamma_1 + \Gamma_2]_\text{min}$ and the following
transcendental equation,
\bel
	y_\text{opt}^{3/2} \approx \frac{8\gamma_\text{LO}\tau N^2}{\sqrt \pi\delta^2
	n_0^\text{opt}} \exp\left[-\frac{1}{y_\text{opt}}\right],
\eel
for $y_\text{opt}$.
The asymptotic solution is 
\bal
	y_\text{opt} &=& \left[\log\left(\frac{8\gamma_\text{LO}\tau N^2}{\sqrt \pi\delta^2
	n_0^\text{opt}}\right)\right]^{-1}\!\!\!\!\!\! \sim
	\left[\log\left(\gamma_\text{LO}\tau N^2\right)\right]^{-1}\quad
	\\
	\label{eq:T_op_GHZ 2}
	T_\text{opt} &=& \frac{\pi^2}{2}\frac{y_\text{opt}}{\gamma_\text{LO}} \sim
	\frac{\pi^2}{2}\frac{[\log(\gamma_\text{LO}\tau N^2)]^{-1}}{\gamma_\text{LO}}
\eal
in the realistic limit of $\gamma_\text{LO}\tau N^2 \gg 1$. The corresponding
minimal value of $\Gamma_1 + \Gamma_2 + \Gamma_3$ is
\begin{align}
	\Big[[\Gamma_1 + &\Gamma_2]_\text{min} + \Gamma_3\Big]_\text{min} \approx \Gamma_1(x_\text{opt},y_\text{opt}) =\frac{n_0^\text{opt}\delta^2}{N^2T_\text{opt}} \\\label{eq:EK4}
	&\sim
	 \gamma_\text{LO}\frac{4\delta^2
	D^2}{\pi^4}\frac{\log(\gamma_\text{LO}\tau N^2)\log(N^2)}{N^2}.
\end{align}

Since $\Gamma_3$ grows exponentially with $T$, interrogation times exceeding $T_\text{opt}$ drastically reduce the resulting laser stability. For averaging times $\tau>T_\text{opt}$ this limit on the maximal interrogation time imposed by phase slip errors, leads to sub-optimal values of the ADEV $\sigma_y(\tau)\propto 1/\sqrt \tau$ according to \refeq{eq:EK9} \& (\ref{eq:EK4}).
However, this limitation can be overcome as we demonstrate in the following section.

If the averaging time $\tau$ is shorter than the
interrogation time suggested by \refeq{eq:T_op_GHZ 2}, $\Gamma_3$
is negligible compared to $\Gamma_1$, and the corresponding effective
linewidth is 
\bel
\label{eq:EK7}
	[\Gamma_1 + \Gamma_2]_\text{min} + \Gamma_3 \approx\frac{n_0^\text{opt}\delta^2}{N^2T} =  \frac{2}{\pi^2}\frac{\delta^2
	D^2}{N^2 T} \log\left(N^2\right),
\eel
and the real optimum is at
$T = \tau$.
The resulting $\tau^{-1}$ scaling indicates that this is the noise-free measurement
regime, and results in a Heisenberg-limited ADEV (up to the logarithmic correction arising from $n_0^\text{opt}$).

\subsection{Extending the Ramsey time beyond the laser noise limit}
\label{sec:BLNL}

As we have seen in the previous Section, for long $\tau>T_\text{opt}$, the cascaded GHZ scheme is limited by the LO
linewidth $\gamma_\text{LO}$ yielding a sub-optimal laser stability [\refeq{eq:EK4}]. In the following we demonstrate a method to efficiently circumvent this problem, by employing additional classical interrogations with varying (effective) interrogation times \cite{Rosenband2013, Borregaard2013}. This allows us to directly assess the digits $Z_0, Z_{-1}, Z_{-2} \dots$, thus effectively countering the problem of phase slips on this level. As such it represents a direct extension of the cascaded GHZ states scheme in the classical domain. 

We assume we have additional $M^*$ groups of $n_0^*$ particles at our disposal.
Using dynamical decoupling techniques \cite{ddc}, we realize that each ensemble
($j=-1,-2\hdots,-M^*$) during the interrogation picks up a phase $\Phi_{j}=
D^{j} \Phi_\text{LO} \mod [-\pi,\pi]$ (alternatively, this can be achieved by
choosing varying interrogation times for each ensemble according to $T_{j} =
D^jT$, for $j<0$ \cite{Rosenband2013}).
This implies that these ensembles evolve successively slower for decreasing $j$,
and thus, in the spirit of 
the section on cascaded GHZ, directly assess the digits
left from the point in the D-numeral representation of $\Phi_\text{LO}$ [compare
\refeq{eq:D-digits}] according to \refeq{eq:Z_j_measuring}, where we now allow
negative values of $j$.

If all digits are correctly estimated, this accounts for all phase slips up to
the last ensemble $j=-M^*$. One readily shows in an analogous calculation to the
one  in 
the section on phase slip errors
that for such a procedure with $M^*$ classical
stages the optimal Ramsey time (i.e., the optimized interrogation time of the
GHZ states) is exponentially prolonged
\bel
\label{eq:EK5}
T^{(M^*)}_\text{opt} = D^{M^*} T_\text{opt}.
\eel
Note, that here we assumed that the total number of particles employed in the
classical part of the scheme is negligible with regard to the total number of
particles, $N^*=M^* n_0^* \ll N$. This is a well justified assumption, as in
order to prolong the optimal Ramsey time by a factor of $k$ from the original
optimum $T_\text{opt}$
 we need a logarithmic number of groups only, $M^*\approx\log_D(k$), as implied
 by \refeq{eq:EK5}.
Furthermore, following the argumentation outlined in 
the section on rounding errors,
we
find that only
\bel
 n_0^* \geq \frac{2}{\pi^2} D^2 \log\left(k N^2\right),
\eel
particles per level are sufficient to ensure that the rounding errors induced by the classical part of the cascade ($j<0$) are negligible. 

 As seen in the previous Section, when the optimal Ramsey time exceeds the averaging time, $T^{(M^*)}_\text{opt} \geq \tau$ the effective linewidth is given by \refeq{eq:EK7} (assuming $N^*\ll N$), as we can neglect the phase slips contribution to the measurement uncertainty ($\Gamma_3$). Extending the Ramsey time to its then optimal (i.e., maximal)  value $T\sim\tau$ we find the ADEV [compare \refeq{eq:EK9}] 
\begin{align}
	\label{eq:sigma_y^2 boxed}
	[\sigma_y(\tau)]_\text{min}&\approx \frac{1}{\omega_0} \frac{\delta\sqrt{ n_0^\text{opt}} }{N\tau}
	\\&\approx \frac{\sqrt 2}{\sqrt \pi\omega_0} \frac{ D\delta}{N\tau} \sqrt{\log\left(N^2\right)}.
\end{align}
This result illustrates that the presented clock protocol achieves Heisenberg-limited clock stability up to a logarithmic correction arising from the number of atoms necessary to compensate for rounding and phase slip errors. The number of particles needed for the extension of the Ramsey time beyond the laser noise limit $T_\text{opt}\approx\gamma_\text{LO}^{-1}$ is given as $N^{*}\approx \frac{2}{\pi^2} D^2 \log\left(k N^2\right)\log_D(k)$ and thus negligible compared to the total particle number $N$. For the optimal choice of basis $D=2$ the constant factor reduces to $ D\delta = 4$. 

The above procedure of interrogation with varying Ramsey times (for the groups $j<0$)  can be understood as a classical pre-narrowing of the laser linewidth \cite{Rosenband2013} to a value that eliminates the threat of phase slips,  before application of the quantum protocol. 

\subsection{Individual qubit noise and final result}
Up to this point we have neglected individual particle dephasing. 
However, as we increase the Ramsey time beyond the laser noise limit $T > \gamma_\text{LO}^{-1}$ we need to consider their effect.

In general, the clock atoms are subject to individual decoherence processes
 characterized by the atomic linewidth
$\gamma_\text{ind}$ ($\ll \gamma_\text{LO}$). 
For  the group with the largest GHZ states in our scheme this leads to an uncertainty contribution of $\ev{\Delta\Phi^2_{M-1}}_\text{dephasing} =
D^{M-1}\gamma_\text{ind}T/n_0$ which results in the measurement uncertainty
\bel
	\label{eq:Dephasing_GHZ 2}
	\ev{\Delta\Phi^2_\text{LO}}_\text{dephasing} =
	\frac{\gamma_\text{ind}T}{n_0 D^{M-1}} \approx
	\frac{\delta\gamma_\text{ind}T}{N},
\eel
which represents a fundamental noise floor in the form of the effective
linewidth contribution $\Gamma_4 =
\ev{\Delta\Phi_\text{LO}^2}_\text{dephasing}/T$.

By adding \refeq{eq:EK7} and $\Gamma_4$, we obtain an approximation of the total
ADEV under single particle noise,
\bal
\label{eq:EK8}
	&&[\sigma_y(\tau)]_\text{min} \approx
	\frac{\delta}{\omega_0 \sqrt{\tau N}}
	\left[
	\frac{1}{T D^{M-1}}
	 +  \gamma_\text{ind}\right]^{1/2}
	 \label{eq:sigma_y boxed GHZ 2},
\eal
where we used $n_0=N/\delta D^{M-1}$.
This equation suggests that the quantum gain in the estimation becomes lost if $T\sim(\gamma_\text{ind}D^{M-1} )^{-1}$. This is a well known result \cite{Huelga1997}, and in fact represents a fundamental limitation of the maximal Ramsey time allowed in the presence of single particle noise \cite{Escher:2011fn}.
We approximate the crossover between the regimes $\tau<(\gamma_\text{ind}D^{M-1} )^{-1}$ ($\tau>(\gamma_\text{ind}D^{M-1} )^{-1}$), where \refeq{eq:EK8} is dominated by the first (second term) by taking $T=\tau$, and rewrite $D^{M-1}$ in terms of the total particle number $N$ to arrive at the final equation characterizing the stability of the cascaded GHZ scheme
\begin{align}\nonumber
	[\sigma_y(\tau)&]_\text{min} \\
	&\approx
	\frac{1}{\omega_0 \sqrt{\tau N}}
	\left[
	\frac{1}{\tau N}\frac{2\delta^2 D^2}{\pi^2}  \log( N^2) +
	  \delta \gamma_\text{ind}\right]^{1/2}.
	 \label{eq:sigma_y boxed GHZ 3}
\end{align}
Again, for the choice of a binary basis $D=2$ the constant factor is given as $\delta D=4$. 

In summary, we find that the cascaded GHZ scheme enables an optimal, Heisenberg-limited laser stability for short averaging times $\tau$ up to a logarithmic correction. This stability reaches the fundamental noise floor given by the single particle dephasing for averaging times $\tau_0 \approx (\gamma_\text{ind}D^{M-1} )^{-1}$. Note, that in the limit $N\rightarrow \infty$ this crossover value goes to zero, and the clock stability is given by the best possible stability allowed by quantum mechanics for all $\tau$. In comparison, classical protocols reach this fundamental limit in the best case \cite{Rosenband2013} at the fixed time $\tau \approx \gamma_\text{ind}^{-1}$. For averaging times larger than this value $\tau\geq \gamma_\text{ind}^{-1}$ the quantum protocol does not offer an advantage over an optimal classical protocol due to fundamental quantum metrological bounds \cite{Escher:2011fn}. 

\section{Analytic solution of $x^n = A\exp[-1/x]$}
\label{sec:Transcendental eq}
To carry out direct optimization of the Allan variance, we need to be able to
solve transcendental equations of the following form
\bel
	\label{eq:Transcendental_eq}
	x^n = A\exp\left[-\frac{1}{x}\right].
\eel
In this Section we obtain an analytic formula for the solution of this equation
over the domain $x\in [0,\infty)$, in the limiting case of $A \gg 1$, where $n$
is real. The sign of $n$ determines the number of solutions: In case of $n>0$,
there are three solutions: $x_{s,0} = 0$, $x_{s,1}\ll 1$ and $x_{s,2}\gg 1$. In
case of $n\leq 0$, there is always a single solution, $x_{s,1} \ll 1$. We are
going to focus on the $x_{s,1} =: x_s$ solution, and give upper and lower
bounds, such that $x_l \leq x_s \leq x_u$, and $x_u/x_l \rightarrow 1$ as $A\rightarrow
\infty$.

The general method of Taylor-expanding the
right side of \refeq{eq:Transcendental_eq}  around $x=0$ fails due to the
non-analytic property of $e^{-1/x}$ function at zero, and forces us to choose an
alternate route. Here, we use a recursion formula, and prove its stability
around $x_s$. Rearranging \refeq{eq:Transcendental_eq} and turning it into
a recursion $f$ yields
\bel
	\label{eq:recursion}
	x_{k} =
	\frac{1}{\log A - n\log x_{k-1}} =: f(x_{k-1}),
\eel 
The iteration of $f$ is stable around the
fixed point ($f(x_s) = x_s$), if and only if $1 > |f'(x_s)| = x_s |n|$,
which is true in the limit $x_s \ll 1$. Stability implies that the fix point can
be obtained as the limit
\bel
	x_s = \lim_{k\rightarrow \infty}x_k = \lim_{k\rightarrow \infty} f^{[k]}(x_0),
\eel
if the $x_0$ starting point is sufficiently close to $x_s$,  where $f^{[k]}$
denotes $k$ iterations of $f$.

In case of $n\leq 0$, $f'(x_s) = n x_s \leq 0$ therefore  $[f^{[k]}(x_0)
- f^{[k-1]}(x_0)]$ is an alternating sequence and we can quickly obtain
upper ($x_u$) and lower ($x_l$) bounds by applying the
recursion $f$ twice on $x_0 = 1$:
\bel
	x_l = x_1 = \frac{1}{\log A} ,\qquad x_u = x_2 = \frac{1}{\log A +
	n\log\log A} ,
\eel

In case of $n>0$ and $x_0 = 1$,  $f^{[k]}(x_s)$
is monotonically decreasing (since $f'(x_s) > 0$), and we can safely
choose the upper bound $x_u = x_2$. To obtain a lower bound, we introduce a new
variable $\xi = - \log x$, and write \refeq{eq:recursion} as
\bal
	\xi_{k} &=& \log\log A + \log\left(1+ \frac{n}{\log A}\xi_{k-1}\right)
	\\
	&\leq&
	\log\log A + \frac{n}{\log A}\xi_{k-1} =: g(\xi_{k-1}),
\eal
where we used that $\log (1+y) \leq y, \; \forall y\in \mathbb{R}^+$ and $g$ is
a new recursion.
Since $g$ is a monotonically increasing function, the  inequality holds for 
multiple iterations, $\xi_{k} \leq g^{[k]}(\xi_0)$, and eventually give the 
upper bound, $-\log x_s = \xi_s \leq \lim_{k\rightarrow \infty}g^{[k]}(\xi_0)$.
In the limit of $A\ll 1$, we can assume $\frac{n}{\log A} < 1$, and the
sequence of iterations of $g$ becomes convergent. Due to its simple form,
we can evaluate its limit in a closed form,  which results in the following
upper bound for $\xi_s$ and the corresponding lower bound for $x_s$.
\bel
	\xi_s < \frac{\log \log A}{1 - \frac{n}{\log
	A}},\qquad\rightarrow \qquad x_s >  \left(\frac{1}{\log
	A}\right)^{\frac{1}{1-\frac{n}{\log A}}}.
\eel
We can obtain an even better (and more conventional) lower bound by applying $f$
once more:
\bel
	x_l = f\left[\left(\frac{1}{\log
	A}\right)^{\frac{1}{1-\frac{n}{\log A}}}\right]	= \frac{1}{\log A +
	\frac{1}{1-\frac{n}{\log A}} n \log\log A}
\eel

For both signs of $n$, $x_l \log A 
\rightarrow 1$, $x_u \log A
\rightarrow 1$, and $x_u/x_l\rightarrow 1$ as $A\rightarrow \infty$, from which
we conclude that
\bel
	\lim_{A\rightarrow \infty}(x_s \log A) = 1.
\eel
For large enough $A$, we can approximate $x_s$ with $[\log A]^{-1}$, and the
relative error is bounded by $|n|\frac{\log\log A}{\log A}$.

\section{Upper bound on the tail of the distribution of the estimated phase}
\label{sec:UpperBound}
The probability of rounding errors is given an upper bound in order to obtain a
more tractable form for optimization. 

\subsection{Upper bound on the tail of the binomial distribution}
\label{sec:Binomial} 
Here we derive an upper bound for the binomial distribution
\bel
	\label{eq:binomial}
	\PP(k) = {n \choose k} p^{k} (1-p)^{n-k}.
\eel
Central limit theorem implies that for large enough $n$, $\PP(k)$ can be
approximated with the normal distribution, $\mathcal{N}(np, np(1-p))$,
however, one can be concerned with the fact that this underestimates the tail of
$\PP(k)$. Here we give $F(k)$ as a strict upper bound on
$\PP(k)$,
\bel
	\label{eq:Upper bound}
	\PP(k) < F(k) = \exp\left[-2(n-1)\left(\frac{k}{n}-p\right)^2\right].
\eel
%
%
%

To see that $F(k)$ is indeed an upper bound of $\PP(k)$ for all $n,p$ and $k$,
let us examine the logarithm of the binomial distribution $\PP(k=ny)$,
\bel
	L(y) = \log{n \choose ny} + ny\log p + n(1-y)\log(1-p),
\eel
where $0\leq y \leq 1$. The properties, we are interested in, are
\begin{itemize}
  \item $L(y)< 0$ for all $y$, since $\PP(k)<1$,
  \item $\left.\frac{\partial}{\partial y}L(y)\right|_{y=p} = 0$,  and positive
  for $y<p$ and negative for $y>p$, since $L(y)=\text{max}$ at $y=p$.
  \item $\frac{\partial^2}{\partial y^2}L(y) = \frac{\partial^2}{\partial
  y^2}{ n \choose ny} = -n^2\Big(\psi_1[1+n(1-y)] + \psi_1[1+ny]\Big)$, where
  $\psi_1(x) = \frac{d}{dx}\log\Gamma(x)$, the first polygamma function.
\end{itemize}
By analyzing the series expansion of $\psi_1(1+\eta)$ for large and small $\eta$
arguments,
\ba
	\psi_1(1+\eta) &=& \frac{1}{\eta} - \frac{1}{2\eta^2} +
	\mathcal{O}(\eta^{-3})\qquad \eta \gg 1\\
	\psi_1(1+\eta) &=& \frac{\pi^2}{6} -2.404\eta + \frac{\pi^4}{30}\eta^2 +
	\mathcal{O}(\eta^3)\qquad \eta \ll 1,
\ea
one can show that $\frac{\partial^2}{\partial y^2}L(y)$ is always
negative and has a global maximum at $y=1/2$, where it takes the value
\bel
	\left.\frac{\partial^2}{\partial y^2}L(y)\right|_{y=1/2} =
	-4(n-1) -
	\mathcal{O}(n^{-1}).
\eel
Therefore the constant function $f''(y) = -4(n-1)$ is an upper bound of $L'' =
\frac{\partial^2}{\partial y^2}L$. Now, let us integrate both
$L''$ and $f''$ twice, and choose the integration constants, so that $L(y) < f(y)$. Since $L'(y=p) = 0$,
\bel
	L'(y) < \intop_p^y\d{\zeta} f''(\zeta) =
	-4(n-1)(y-p),
\eel
which is chosen to be $f'(y)$, and since $L(p) < 0$,
\bel
	L(y) < \intop_p^y \d{\zeta} L'(\zeta) <
	\intop_p^y \d{\zeta} f'(\zeta) = -2(n-1)(y-p)^2,
\eel
which is chosen to be $f(y)$. 
Consequently
\bel
	\PP(ny) = \exp [L(y)]  < \exp [f(y)] = F(ny).
\eel

\subsection{Upper bound on the distribution of the estimated phase}
\label{sec:Phase_estimation}

Here we give an upper bound on the distribution of the Ramsey phase $\Phi$, as
determined by estimating $\cos\Phi$ and $\sin\Phi$ from two sub-groups of
GHZ states ($X,Y$), each providing $n/2$ measurement (parity) bits.
Qubits in group $X$ are prepared in $[\ket{\mathbf{0}} +
\ket{\mathbf{1}}]/\sqrt{2} =:
\ket{+}$, and measured in $\ket{+}$ with probability $p_x = [1+\cos\Phi]/2$
 after time $T$, while qubits in ensemble $Y$ are prepared in $[\ket{\mathbf{0}} +
 i\ket{\mathbf{1}}]/\sqrt{2}$, and measured in $\ket{+}$  with probability $p_y
 = [1+\sin\Phi]/2$ after time $T$. The number of $\ket{+}$ outcomes $k_x$ and
 $k_y$ from groups $X$ and $Y$, respectively are binomial random variables
 with the distribution
 \bel
 	\PP(k_\nu) = {n/2 \choose k_\nu} p_\nu^{k_\nu} (1-p_\nu)^{n/2-k_\nu},
 \eel
where $\nu \in\{ x,y\}$. Using the upper bound from \refeq{eq:Upper bound},
and noting that $n/2 > 1$ we can give the following upper bound on the joint
distribution of $k_x$ and $k_y$,
\bel
	\PP(k_x,k_y) < \exp\left[-\frac{n}{2}\left(\frac{2k_x}{n} - p_x\right)^2
	-\frac{n}{2}\left(\frac{2k_y}{n} - p_y\right)^2\right].
\eel
Let us introduce the polar coordinates $r,\varphi$: $r\cos\varphi =
\frac{2k_x}{n} - \frac{1}{2}$ and $r\sin\varphi = \frac{2k_y}{n} -
\frac{1}{2}$, and smear the distribution into a continuous  density
function $\rho(r,\varphi)$, and its upper bound accordingly:
\bel
	\rho(r,\varphi) < \frac{n^2}{4} r \exp\left[-\frac{n}{2}\left(r^2 -
	2r\cos(\varphi-\Phi^\text{real}) + 1\right)\right].
\eel
The marginal distribution of $r$ is independent of $\Phi^\text{real}$, which
means that $r$ does not hold any information about $\Phi^\text{real}$. The upper
bound on the marginal distribution of $\varphi$ can be written as
\bal
	\rho(\varphi) &<&  \left(\frac{n}{4} +
	\frac{n^{3/2}\sqrt{\pi}}{\sqrt{32}}\right)\exp\left[-\frac{n}{2}\sin^2(\varphi
	- \Phi^\text{real})\right]
	\\
	\label{eq:Tail}
	& \sim &
	n^{3/2} \exp\left[-\frac{n}{2}(\varphi - \Phi^\text{real})^2\right],
\eal
where in the second line we assumed $|\varphi - \Phi^\text{real}| \ll 1$, and
$n\gg 1$. This result is an upper bound on the distribution of $\varphi$, which
we are going to use to give an upper bound on the rounding error probability,
\bel 
	P_\text{re} = 2\intop_{\pi/D}^{\infty}\d{\varphi} \rho(\varphi +
	\Phi^\text{real}).
\eel
The rigorous upper bound on the tail of $\rho$ is provided by \refeq{eq:Tail},
as long as $\pi/D \ll 1$, and $n \gg 1$.


\begin{thebibliography}{22}
\expandafter\ifx\csname natexlab\endcsname\relax\def\natexlab#1{#1}\fi
\expandafter\ifx\csname bibnamefont\endcsname\relax
  \def\bibnamefont#1{#1}\fi
\expandafter\ifx\csname bibfnamefont\endcsname\relax
  \def\bibfnamefont#1{#1}\fi
\expandafter\ifx\csname citenamefont\endcsname\relax
  \def\citenamefont#1{#1}\fi
\expandafter\ifx\csname url\endcsname\relax
  \def\url#1{\texttt{#1}}\fi
\expandafter\ifx\csname urlprefix\endcsname\relax\def\urlprefix{URL }\fi
\providecommand{\bibinfo}[2]{#2}
\providecommand{\eprint}[2][]{\url{#2}}

\bibitem[{\citenamefont{Nicholson et~al.}(2012)\citenamefont{Nicholson, Martin,
  Williams, Bloom, Bishof, Swallows, Campbell, and Ye}}]{Nicholson2012}
\bibinfo{author}{\bibfnamefont{T.~L.} \bibnamefont{Nicholson}},
  \bibinfo{author}{\bibfnamefont{M.~J.} \bibnamefont{Martin}},
  \bibinfo{author}{\bibfnamefont{J.~R.} \bibnamefont{Williams}},
  \bibinfo{author}{\bibfnamefont{B.~J.} \bibnamefont{Bloom}},
  \bibinfo{author}{\bibfnamefont{M.}~\bibnamefont{Bishof}},
  \bibinfo{author}{\bibfnamefont{M.~D.} \bibnamefont{Swallows}},
  \bibinfo{author}{\bibfnamefont{S.~L.} \bibnamefont{Campbell}},
  \bibnamefont{and} \bibinfo{author}{\bibfnamefont{J.}~\bibnamefont{Ye}},
  \bibinfo{journal}{Phys. Rev. Lett.} \textbf{\bibinfo{volume}{109}},
  \bibinfo{pages}{230801} (\bibinfo{year}{2012}).

\bibitem[{\citenamefont{Lemke et~al.}(2009)\citenamefont{Lemke, Ludlow, Barber,
  Fortier, Diddams, Jiang, Jefferts, Heavner, Parker, and Oates}}]{Lemke2009}
\bibinfo{author}{\bibfnamefont{N.}~\bibnamefont{Lemke}},
  \bibinfo{author}{\bibfnamefont{A.}~\bibnamefont{Ludlow}},
  \bibinfo{author}{\bibfnamefont{Z.}~\bibnamefont{Barber}},
  \bibinfo{author}{\bibfnamefont{T.}~\bibnamefont{Fortier}},
  \bibinfo{author}{\bibfnamefont{S.}~\bibnamefont{Diddams}},
  \bibinfo{author}{\bibfnamefont{Y.}~\bibnamefont{Jiang}},
  \bibinfo{author}{\bibfnamefont{S.}~\bibnamefont{Jefferts}},
  \bibinfo{author}{\bibfnamefont{T.}~\bibnamefont{Heavner}},
  \bibinfo{author}{\bibfnamefont{T.}~\bibnamefont{Parker}}, \bibnamefont{and}
  \bibinfo{author}{\bibfnamefont{C.}~\bibnamefont{Oates}},
  \bibinfo{journal}{Phys. Rev. Lett.} \textbf{\bibinfo{volume}{103}},
  \bibinfo{pages}{063001} (\bibinfo{year}{2009}).

\bibitem[{\citenamefont{Chou et~al.}(2010)\citenamefont{Chou, Hume, Koelemeij,
  Wineland, and Rosenband}}]{Chou2010}
\bibinfo{author}{\bibfnamefont{C.~W.} \bibnamefont{Chou}},
  \bibinfo{author}{\bibfnamefont{D.~B.} \bibnamefont{Hume}},
  \bibinfo{author}{\bibfnamefont{J.~C.~J.} \bibnamefont{Koelemeij}},
  \bibinfo{author}{\bibfnamefont{D.~J.} \bibnamefont{Wineland}},
  \bibnamefont{and}
  \bibinfo{author}{\bibfnamefont{T.}~\bibnamefont{Rosenband}},
  \bibinfo{journal}{Phys. Rev. Lett.} \textbf{\bibinfo{volume}{104}},
  \bibinfo{pages}{070802} (\bibinfo{year}{2010}).

\bibitem[{\citenamefont{Bishof et~al.}(2013)\citenamefont{Bishof, Zhang,
  Martin, and Ye}}]{Bloom2013}
\bibinfo{author}{\bibfnamefont{B.~J.}~\bibnamefont{Bloom}} \textit{et al.} \eprint{arXiv:1309.1137} (\bibinfo{year}{2013}).

\bibitem[{\citenamefont{Eckstein et~al.}(1978)\citenamefont{Eckstein, Ferguson,
  and H\"{a}nsch}}]{Eckstein1978}
\bibinfo{author}{\bibfnamefont{J.}~\bibnamefont{Eckstein}},
  \bibinfo{author}{\bibfnamefont{A.}~\bibnamefont{Ferguson}}, \bibnamefont{and}
  \bibinfo{author}{\bibfnamefont{T.}~\bibnamefont{H\"{a}nsch}},
  \bibinfo{journal}{Phys. Rev. Lett.} \textbf{\bibinfo{volume}{40}},
  \bibinfo{pages}{847} (\bibinfo{year}{1978}).

\bibitem[{\citenamefont{Reichert et~al.}(2000)\citenamefont{Reichert, Niering,
  Holzwarth, Weitz, Udem, and Hansch}}]{Reichert2000}
\bibinfo{author}{\bibfnamefont{J.}~\bibnamefont{Reichert}},
  \bibinfo{author}{\bibfnamefont{M.}~\bibnamefont{Niering}},
  \bibinfo{author}{\bibfnamefont{R.}~\bibnamefont{Holzwarth}},
  \bibinfo{author}{\bibfnamefont{M.}~\bibnamefont{Weitz}},
  \bibinfo{author}{\bibfnamefont{T.}~\bibnamefont{Udem}}, \bibnamefont{and}
  \bibinfo{author}{\bibfnamefont{T.}~\bibnamefont{Hansch}},
  \bibinfo{journal}{Phys. Rev. Lett.} \textbf{\bibinfo{volume}{84}},
  \bibinfo{pages}{3232} (\bibinfo{year}{2000}).

\bibitem[{\citenamefont{Jones et~al.}(2000)\citenamefont{Jones, Diddams, Ranka,
  Stentz, Windeler, Hall, and Cundiff}}]{Jones2000}
\bibinfo{author}{\bibfnamefont{D.~J.} \bibnamefont{Jones}},
  \bibinfo{author}{\bibfnamefont{S.~A.} \bibnamefont{Diddams}},
  \bibinfo{author}{\bibfnamefont{J.~K.} \bibnamefont{Ranka}},
  \bibinfo{author}{\bibfnamefont{A.}~\bibnamefont{Stentz}},
  \bibinfo{author}{\bibfnamefont{R.~S.} \bibnamefont{Windeler}},
  \bibinfo{author}{\bibfnamefont{J.~L.} \bibnamefont{Hall}}, \bibnamefont{and}
  \bibinfo{author}{\bibfnamefont{S.~T.} \bibnamefont{Cundiff}},
  \bibinfo{journal}{Science} \textbf{\bibinfo{volume}{288}},
  \bibinfo{pages}{635} (\bibinfo{year}{2000}).

\bibitem[{\citenamefont{Ye et~al.}(2003)\citenamefont{Ye, Peng, Jones, Holman,
  Hall, Jones, Diddams, Kitching, Bize, Bergquist et~al.}}]{Ye2003}
\bibinfo{author}{\bibfnamefont{J.}~\bibnamefont{Ye}},
  \bibinfo{author}{\bibfnamefont{J.-L.} \bibnamefont{Peng}},
  \bibinfo{author}{\bibfnamefont{R.~J.} \bibnamefont{Jones}},
  \bibinfo{author}{\bibfnamefont{K.~W.} \bibnamefont{Holman}},
  \bibinfo{author}{\bibfnamefont{J.~L.} \bibnamefont{Hall}},
  \bibinfo{author}{\bibfnamefont{D.~J.} \bibnamefont{Jones}},
  \bibinfo{author}{\bibfnamefont{S.~a.} \bibnamefont{Diddams}},
  \bibinfo{author}{\bibfnamefont{J.}~\bibnamefont{Kitching}},
  \bibinfo{author}{\bibfnamefont{S.}~\bibnamefont{Bize}},
  \bibinfo{author}{\bibfnamefont{J.~C.} \bibnamefont{Bergquist}},
  \bibnamefont{et~al.}, \bibinfo{journal}{J. Opt. Soc. Am. B}
  \textbf{\bibinfo{volume}{20}}, \bibinfo{pages}{1459} (\bibinfo{year}{2003}).

\bibitem[{\citenamefont{Bu\v{z}ek et~al.}(1999)\citenamefont{Bu\v{z}ek, Derka,
  and Massar}}]{Buzek1999}
\bibinfo{author}{\bibfnamefont{V.}~\bibnamefont{Bu\v{z}ek}},
  \bibinfo{author}{\bibfnamefont{R.}~\bibnamefont{Derka}}, \bibnamefont{and}
  \bibinfo{author}{\bibfnamefont{S.}~\bibnamefont{Massar}},
  \bibinfo{journal}{Phys. Rev. Lett.} \textbf{\bibinfo{volume}{82}},
  \bibinfo{pages}{2207} (\bibinfo{year}{1999}).

\bibitem[{\citenamefont{Andr\'{e} et~al.}(2004)\citenamefont{Andr\'{e},
  S{\o}rensen, and Lukin}}]{Andre2004}
\bibinfo{author}{\bibfnamefont{A.}~\bibnamefont{Andr\'{e}}},
  \bibinfo{author}{\bibfnamefont{A.}~\bibnamefont{S{\o}rensen}},
  \bibnamefont{and} \bibinfo{author}{\bibfnamefont{M.}~\bibnamefont{Lukin}},
  \bibinfo{journal}{Phys. Rev. Lett.} \textbf{\bibinfo{volume}{92}},
  \bibinfo{pages}{230801} (\bibinfo{year}{2004}).
  
      
  \bibitem[{\citenamefont{Chauvet}(2010)}]{LouchetChauvet:2010fs}
\bibinfo{author}{\bibfnamefont{A.}~\bibnamefont{Louchet-Chauvet}} \bibnamefont{et~al.}, 
\bibinfo{journal}{New J. Phys.} \textbf{\bibinfo{volume}{12}},
  \bibinfo{pages}{065032} (\bibinfo{year}{2010}).

\bibitem[{\citenamefont{Rosenband}(2012)}]{Rosenband2012_numerical}
\bibinfo{author}{\bibfnamefont{T.}~\bibnamefont{Rosenband}}, \eprint{arXiv:1203.0288} 
  (\bibinfo{year}{2012}).
  
  

\bibitem[{\citenamefont{Borregaard and
  S{\o}rensen}(2013{\natexlab{a}})}]{Borregaard2013_nearHeisenberg}
\bibinfo{author}{\bibfnamefont{J.}~\bibnamefont{Borregaard}} \bibnamefont{and}
  \bibinfo{author}{\bibfnamefont{A.~S.} \bibnamefont{S{\o}rensen}}, \bibinfo{journal}{Phys. Rev. Lett.} \textbf{\bibinfo{volume}{111}},
  \bibinfo{pages}{090801} (\bibinfo{year}{2013}).
  
    
  \bibitem[{\citenamefont{Huelga}(1997)}]{Huelga1997}
\bibinfo{author}{\bibfnamefont{S.}~\bibnamefont{Huelga}} \bibnamefont{et~al.}, 
\bibinfo{journal}{Phys. Rev. Lett.} \textbf{\bibinfo{volume}{79}},
  \bibinfo{pages}{3865} (\bibinfo{year}{1997}).
  
  
\bibitem[{\citenamefont{Bollinger et~al.}(1996)\citenamefont{Bollinger, Itano,
  Wineland, and Heinzen}}]{Bollinger1996}
\bibinfo{author}{\bibfnamefont{J.}~\bibnamefont{Bollinger}},
  \bibinfo{author}{\bibfnamefont{W.}~\bibnamefont{Itano}},
  \bibinfo{author}{\bibfnamefont{D.}~\bibnamefont{Wineland}}, \bibnamefont{and}
  \bibinfo{author}{\bibfnamefont{D.}~\bibnamefont{Heinzen}},
  \bibinfo{journal}{Phys. Rev. A} \textbf{\bibinfo{volume}{54}},
  \bibinfo{pages}{R4649} (\bibinfo{year}{1996}).
  
  \bibitem[{\citenamefont{Wineland et~al.}(1998)\citenamefont{Wineland, Monroe,
  Itano, Leibfried, King, and Meekhof}}]{Wineland1998}
\bibinfo{author}{\bibfnamefont{D.}~\bibnamefont{Wineland}},
  \bibinfo{author}{\bibfnamefont{C.}~\bibnamefont{Monroe}},
  \bibinfo{author}{\bibfnamefont{W.}~\bibnamefont{Itano}},
  \bibinfo{author}{\bibfnamefont{D.}~\bibnamefont{Leibfried}},
  \bibinfo{author}{\bibfnamefont{B.}~\bibnamefont{King}}, \bibnamefont{and}
  \bibinfo{author}{\bibfnamefont{D.}~\bibnamefont{Meekhof}},
  \bibinfo{journal}{J. Res. Natl. Inst. Stan.} \textbf{\bibinfo{volume}{103}},
  \bibinfo{pages}{259} (\bibinfo{year}{1998}).
  
     \bibitem[{\citenamefont{Rosenband and Leibrandt}(2013)}]{Rosenband2013}
\bibinfo{author}{\bibfnamefont{T.}~\bibnamefont{Rosenband}} \bibnamefont{and}
  \bibinfo{author}{\bibfnamefont{D.~R.} \bibnamefont{Leibrandt}}, \eprint{arXiv:1303.6357} (\bibinfo{year}{2013}).

\bibitem[{\citenamefont{Borregaard and
  S{\o}rensen}(2013{\natexlab{b}})}]{Borregaard2013}
\bibinfo{author}{\bibfnamefont{J.}~\bibnamefont{Borregaard}} \bibnamefont{and}
  \bibinfo{author}{\bibfnamefont{A.~S.} \bibnamefont{S{\o}rensen}},
  \bibinfo{journal}{Phys. Rev. Lett.} \textbf{\bibinfo{volume}{111}},
  \bibinfo{pages}{090802} (\bibinfo{year}{2013}).

  
 \bibitem[{\citenamefont{Nielsen and Chuang}(2011)}]{Nielsen_Chuang}
\bibinfo{author}{\bibfnamefont{M.~A.} \bibnamefont{Nielsen}} \bibnamefont{and}
  \bibinfo{author}{\bibfnamefont{I.~L.} \bibnamefont{Chuang}},
  \emph{\bibinfo{title}{Quantum Computation and Quantum Information}}
  (\bibinfo{publisher}{Cambridge University Press}, \bibinfo{year}{2011}).
  
     
      \bibitem[{\citenamefont{Giovanetti et~al.}(1993)}]{Giovanetti2011}
\bibinfo{author}{\bibfnamefont{V.}~\bibnamefont{Giovanetti}},
  \bibinfo{author}{\bibfnamefont{S.}~\bibnamefont{Lloyd}}, \bibnamefont{and}
  \bibinfo{author}{\bibfnamefont{L.}~\bibnamefont{Maccone}},
  \bibinfo{journal}{Nature Photonics} \textbf{\bibinfo{volume}{5}},
  \bibinfo{pages}{222} (\bibinfo{year}{2011}).
  
  \bibitem [{\citenamefont {Uhrig}\ \emph {et~al.}(1997)\citenamefont {Huelga},
  \citenamefont {Macchiavello}, \citenamefont {Pellizzari}, \citenamefont
  {Ekert}, \citenamefont {Plenio},\ and\ \citenamefont {Cirac}}]{ddc}%
  \BibitemOpen
  \bibfield  {author} {\bibinfo {author} {\bibfnamefont {G.~S.}~\bibnamefont
  {Uhrig}},\ }
  {\bibfield  {journal} {\bibinfo  {journal} {Phys. Rev. Lett.}\
  }\textbf {\bibinfo {volume} {98}},\ \bibinfo {pages} {100504} (\bibinfo {year}
  {2007})}.%


\bibitem[{\citenamefont{Andr\'{e}}(2005)}]{fn1}
Alternatively,  it is also possible to perform
direct phase feedback.

\bibitem[{\citenamefont{Andr\'{e}}(2005)}]{SI}
See Supplemental Information for details.

\bibitem[{\citenamefont{Andr\'{e}}(2005)}]{Andre2005}
\bibinfo{author}{\bibfnamefont{A.}~\bibnamefont{Andr\'{e}}}, Ph.D. thesis,
  \bibinfo{school}{Harvard Univeristy, Cambridge, Massachusetts}
  (\bibinfo{year}{2005}).

\bibitem[{\citenamefont{Santarelli et~al.}(1998)\citenamefont{Santarelli,
  Audoin, Makdissi, Laurent, Dick, and Clairon}}]{Santarelli1998}
\bibinfo{author}{\bibfnamefont{G.}~\bibnamefont{Santarelli}},
  \bibinfo{author}{\bibfnamefont{C.}~\bibnamefont{Audoin}},
  \bibinfo{author}{\bibfnamefont{A.}~\bibnamefont{Makdissi}},
  \bibinfo{author}{\bibfnamefont{P.}~\bibnamefont{Laurent}},
  \bibinfo{author}{\bibfnamefont{G.~J.} \bibnamefont{Dick}}, \bibnamefont{and}
  \bibinfo{author}{\bibfnamefont{A.}~\bibnamefont{Clairon}},
  \bibinfo{journal}{IEEE Trans. Ultrason. Ferroelectr. Freq. Control} \textbf{\bibinfo{volume}{45}},
  \bibinfo{pages}{887} (\bibinfo{year}{1998}).

\bibitem[{\citenamefont{Rao}(1945)}]{Rao1945}
\bibinfo{author}{\bibfnamefont{H.} \bibnamefont{Cram\'er}},
  \emph{\bibinfo{title}{Mathematical Methods of Statistics}}
  (\bibinfo{publisher}{Princeton Univ.}, \bibinfo{year}{1946}).



\bibitem[{\citenamefont{Itano et~al.}(1993)\citenamefont{Itano, Bergquist,
  Bollinger, Gilligan, Heinzen, Moore, Raizen, and Wineland}}]{Itano1993}
\bibinfo{author}{\bibfnamefont{W.}~\bibnamefont{Itano}},
  \bibinfo{author}{\bibfnamefont{J.}~\bibnamefont{Bergquist}},
  \bibinfo{author}{\bibfnamefont{J.}~\bibnamefont{Bollinger}},
  \bibinfo{author}{\bibfnamefont{J.}~\bibnamefont{Gilligan}},
  \bibinfo{author}{\bibfnamefont{D.}~\bibnamefont{Heinzen}},
  \bibinfo{author}{\bibfnamefont{F.}~\bibnamefont{Moore}},
  \bibinfo{author}{\bibfnamefont{M.}~\bibnamefont{Raizen}}, \bibnamefont{and}
  \bibinfo{author}{\bibfnamefont{D.}~\bibnamefont{Wineland}},
  \bibinfo{journal}{Phys. Rev. A} \textbf{\bibinfo{volume}{47}},
  \bibinfo{pages}{3554} (\bibinfo{year}{1993}).
  
\bibitem[{\citenamefont{Andr\'{e}}(2005)}]{fn2}
  $\Phi_j$ are
estimated within the entire $[-\pi, \pi)$ interval by using pairs of
sub-ensembles to measure $\cos\Phi_j$ and $\sin\Phi_j$
simultaneously \cite{Rosenband2013}.
  
    \bibitem[{\citenamefont{Giedke et~al.}(1993)}]{Geza}
\bibinfo{author}{\bibfnamefont{G.}~\bibnamefont{Giedke}},
  \bibinfo{author}{\bibfnamefont{J.~M.}~\bibnamefont{Taylor}},
   \bibinfo{author}{\bibfnamefont{D.}~\bibnamefont{D'Alessandro}},
    \bibinfo{author}{\bibfnamefont{M.~D.}~\bibnamefont{Lukin}},
   \bibnamefont{and}
  \bibinfo{author}{\bibfnamefont{A.}~\bibnamefont{Imamo\u{g}lu}},
  \bibinfo{journal}{Phys. Rev. A} \textbf{\bibinfo{volume}{74}},
  \bibinfo{pages}{032316} (\bibinfo{year}{2006}).


\bibitem[{\citenamefont{Escher et~al.}(2011)\citenamefont{Escher,
  de~Matos~Filho, and Davidovich}}]{Escher:2011fn}
\bibinfo{author}{\bibfnamefont{B.~M.} \bibnamefont{Escher}},
  \bibinfo{author}{\bibfnamefont{R.~L.} \bibnamefont{de~Matos~Filho}},
  \bibnamefont{and}
  \bibinfo{author}{\bibfnamefont{L.}~\bibnamefont{Davidovich}},
  \bibinfo{journal}{Nature Physics} \textbf{\bibinfo{volume}{7}},
  \bibinfo{pages}{406} (\bibinfo{year}{2011}).
  
  
    \bibitem[{\citenamefont{Meiser}(2008)}]{Meiser:2008fo}
\bibinfo{author}{\bibfnamefont{D.}~\bibnamefont{Meiser}},
\bibinfo{author}{\bibfnamefont{J.}~\bibnamefont{Ye}},
\bibinfo{author}{\bibfnamefont{M.~J.}~\bibnamefont{Holland}}. 
\bibinfo{journal}{New J. Phys.} \textbf{\bibinfo{volume}{10}},
  \bibinfo{pages}{073014} (\bibinfo{year}{2008}).
   
  
  \bibitem[{\citenamefont{Komar et~al.}(2013)}]{tbp}
\bibinfo{author}{\bibfnamefont{P.} \bibnamefont{K\'om\'ar}},
  \bibinfo{author}{\bibfnamefont{E.~M.} \bibnamefont{Kessler}},
  \bibnamefont{\textit{et al.}}
  \bibinfo{journal}{(in preparation)}.

\end{thebibliography}

\end{document}